\documentclass[10pt,letterpaper]{article}
\usepackage[]{fontenc}
\usepackage[latin9]{inputenc}
\usepackage[letterpaper]{geometry}
\usepackage{xcolor}
\usepackage{pdfcolmk}
\usepackage{prettyref}
\usepackage{amsmath}
\usepackage{graphicx}
\usepackage{setspace}
\usepackage{amssymb}
\PassOptionsToPackage{normalem}{ulem}
\usepackage{ulem}
\usepackage[unicode=true, 
 bookmarks=true,bookmarksnumbered=true,bookmarksopen=true,bookmarksopenlevel=3,
 breaklinks=false,pdfborder={0 0 0},backref=false,colorlinks=false]
 {hyperref}
\hypersetup{pdftitle={Quantum Strings and the AdS4/CFT3 Interpolating Function},
 pdfauthor={Michael C Abbott, Ines Aniceto, Diego Bombardelli}}
 
\makeatletter

\providecommand{\tabularnewline}{\\}
\providecolor{lyxadded}{rgb}{0,0,1}
\providecolor{lyxdeleted}{rgb}{1,0,0}

\usepackage[T1]{fontenc}
\usepackage{ae}
\usepackage{aecompl}

\usepackage[english]{babel}
\usepackage{datetime}
\shortdate 

\usepackage[sf,bf,small]{titlesec}

\usepackage[small,sf,bf,hang]{caption}

\usepackage[multiple]{footmisc}
\long\def\symbolfootnote[#1]#2{\begingroup%
\def\thefootnote{\fnsymbol{footnote}}\footnote[#1]{#2}\endgroup}

\usepackage{titletoc}
\dottedcontents{section}[6mm]{\smallskip }{6mm}{0pc} 
\titlecontents*{subsection}[6mm]{\itshape\small }{\thecontentslabel \ }{}{, \thecontentspage}[ \ \textbullet \ \ ][] 
\def\tableofcontents{\subsection*{\contentsname}\vspace{-2mm}\@starttoc{toc}}

\usepackage[nosort]{cite} 

\DeclareMathOperator{\eig}{eig}

\AtBeginDocument{
  
}

\makeatother

\begin{document}
\begin{flushright}
{\small arXiv:1006.2174}\\
{\small TIFR/TH/10-15\bigskip}
\par\end{flushright}{\small \par}

\begin{center}
\textsf{\textbf{\Large Quantum Strings and the}} \textsf{\textbf{\Large \smallskip\smallskip\smallskip
}}\\
\textsf{\textbf{\Large AdS$_{\mathsf{4}}$/CFT$_{\mathsf{3}}$
Interpolating Function}}\vspace{-3mm}
\par\end{center}

\begin{singlespace}
\begin{center}
Michael C. Abbott$,^{1}$ In\^{e}s Aniceto$^{\,2}$ and Diego Bombardelli$^{\,3}$\emph{
}\bigskip \\
{\small $^{1}$}\emph{\small{} Tata Institute of Fundamental Research,
}\\
\emph{\small Homi Bhabha Rd, Mumbai 400-005, India}\\
\emph{\small abbott@theory.tifr.res.in }{\small \bigskip }\emph{\small }\\
{\small $^{2}$}\emph{\small{} CAMGSD, Departamento de Matem\'{a}tica,
Instituto Superior T\'{e}cnico, }\\
\emph{\small Av. Rovisco Pais, 1049-001 Lisboa, Portugal}{\small }\\
\emph{\small ianiceto@math.ist.utl.pt}{\small \bigskip }\emph{\small }\\
{\small $^{3}$}\emph{\small{} Dipartimento di Fisica, Universit\`{a}
di Bologna,}{\small{} }\emph{\small }\\
\emph{\small{} Via Irnerio 46, 40126 Bologna, Italy}{\small }\\
\emph{\small \&}{\small }\\
\emph{\small Department of Physics and Institute for the Early
Universe }\\
\emph{\small Ewha Womans University, DaeHyun 11-1, Seoul 120-750,
South Korea}{\small }\\
\emph{\small diegobombardelli@gmail.com}
\par\end{center}{\small \par}
\end{singlespace}

\begin{center}
\bigskip 11 June 2010\vspace{-3mm}
\par\end{center}

\subsection*{\hspace{9mm}Abstract}
\begin{quote}
The existence of a nontrivial interpolating function\textsf{\textbf{\Large{}
}}$h(\lambda)$ is one of the novel features of the new AdS$_{4}$/CFT$_{3}$
correspondence involving ABJM theory. At strong coupling, most of
the investigation of semiclassical effects so far has been for strings
in the $AdS_{4}$ sector. Several cutoff prescriptions have been proposed,
leading to different predictions for the constant term in the expansion
$h(\lambda)=\sqrt{\lambda/2}+c+\ldots$. We calculate quantum corrections
for giant magnons, using the algebraic curve, and show by comparing
to the dispersion relation that the same prescriptions lead to the
same values of $c$ in this $CP^{3}$ sector. We then turn to finite-$J$
effects, where a comparison with the L\"{u}scher F-term correction
shows a mismatch for one of the three sum prescriptions. We also compute
some dyonic and higher F-terms for future comparisons. \thispagestyle{empty} 
\end{quote}
\tableofcontents{}

\section{Introduction}

In the AdS/CFT correspondence \cite{Maldacena:1997re,Aharony:2008ug}
between ABJM's superconformal Chern--Simons-matter theory and IIA
strings on $AdS_{4}\times CP^{3}$, the dispersion relation for a
bound state of $Q$ magnons (or a dyonic giant magnon) is \begin{equation}
E\equiv\Delta-\frac{J}{2}=\sqrt{\frac{Q^{2}}{4}+4h(\lambda)^{2}\sin^{2}\frac{p}{2}\:}.\label{eq:disp-rel-ABJM}\end{equation}
One important difference from the $AdS_{5}\times S^{5}$ example is
that $h(\lambda)$ is now a nontrivial function. It is related to
the 't Hooft coupling $\lambda$ as follows:%
\footnote{We will often use $g=\sqrt{\lambda/8}$ instead of $\lambda$, matching
the conventions of \cite{Gromov:2008bz,Abbott:2009um,Bandres:2009kw},
and also $\alpha=\Delta/2g$.%
}\begin{align}
h(\lambda) & =\sqrt{\dfrac{\lambda}{2}}+c+\mathcal{O}\Big(\frac{1}{\sqrt{\lambda}}\Big),\qquad\lambda\gg1\label{eq:two-expansions-of-h}\\
h(\lambda)^{2} & =\lambda^{2}+h_{4}\lambda^{4}+\mathcal{O}(\lambda^{6}),\qquad\:\lambda\ll1.\nonumber \end{align}
The leading terms here come from the comparison with classical strings
and with two-loop gauge-theory results \cite{Aharony:2008ug,Gaiotto:2008cg,Grignani:2008is,Nishioka:2008gz,Minahan:2008hf,Minahan:2009aq}.
Both sides involve $\lambda$ via the AdS/CFT relation%
\footnote{As usual, $R$ is the radius of $CP^{3}$ and $\sqrt{\alpha'}$ the
string scale, $N$ the rank of the gauge group and $k$ the level
number. And (on the string side) $\Delta$ is the energy, $J$ and
$Q$ are $CP^{3}$ angular momenta, and $S$ is an $AdS_{4}$ angular
momentum.%
}\begin{equation}
\frac{R^{4}}{2^{5}\pi^{2}\alpha^{\prime2}}=\lambda=\frac{N}{k}.\label{eq:relation-R4-lambda-N/k}\end{equation}
Four-loop gauge theory calculations \cite{Minahan:2009aq,Minahan:2009wg}
show that $h_{4}=-4\zeta(2)\approx-6.58$.%
\footnote{Versions of \cite{Minahan:2009aq,Minahan:2009wg} before October 2010
gave instead $h_{4}=-16+4\zeta(2)\approx-9.42$.%
} This rules out various simple interpolating functions one could imagine
from the leading behaviours \cite{Nishioka:2008gz,Gromov:2008qe}. 

The main goal of this paper is to calculate the value of $c$ from
the one-loop corrections to the dispersion relation \eqref{eq:disp-rel-ABJM}
of the giant magnon. The next subsection of the introduction reviews
some previous calculations of $c$, which used a different classical
solution. After that we discuss the cutoff prescriptions used, before
turning to giant magnons in section \ref{sub:Giant-magnons-intro}.

\subsection[Spinning strings in AdS]{The coefficient $c$ from spinning strings in AdS\label{sub:Spinning-strings-in-AdS}}

A number of early ABJM papers studied spinning strings in an $AdS_{3}$
subspace. These have \cite{Gubser:2002tv} \begin{equation}
\Delta-S=f(\lambda)\log S\label{eq:AdS-string-classical}\end{equation}
and at leading order $f(\lambda)=\sqrt{2\lambda}$ \cite{Aharony:2008ug}.
Two very different calculations of the one-loop, $o(1)$, semiclassical
corrections were done:
\begin{itemize}
\item Several authors \cite{McLoughlin:2008ms,Alday:2008ut,Krishnan:2008zs}
found explicit modes using the worldsheet action, and obtained\[
\delta E_{\mathrm{old}}=-5\frac{\log2}{2\pi}\log S.\]
Despite the classical solution being identical to those studied in
$AdS_{5}\times S^{5}$, this quantum result is different to that of
\cite{Frolov:2002av,Park:2005ji,Frolov:2006qe}. And the logic is
that small fluctuations explore not only the $AdS_{3}$ subspace,
but the other directions too.
\item Using the proposed all-loop $sl(2)$ Bethe ansatz, \cite{Gromov:2008qe}
obtained\[
\delta E_{\mathrm{BA}}=-3\frac{\log2}{2\pi}\log S.\]
Apart from trivial changes of constants, and one minus, the Bethe
equations used for this $sl(2)$ sector are identical to those used
in the $AdS_{5}\times S^{5}$ case \cite{Gromov:2008fy,Arutyunov:2004vx,Frolov:2006qe}.
\end{itemize}
Two ways to resolve this apparent discrepancy have been proposed.
One is to notice that while the string calculation is an expansion
in $1/\sqrt{\lambda}$, the Bethe ansatz calculation is a series in
$1/h(\lambda)$. Expanding the latter in $1/\sqrt{\lambda}$ we can
compare them: \begin{align*}
f(\lambda) & =2h(\lambda)-3\frac{\log2}{2\pi}+\mathcal{O}\Big(\frac{1}{h}\Big)\\
 & =\sqrt{2\lambda}+\left(2c-3\frac{\log2}{2\pi}\right)+\mathcal{O}\Big(\frac{1}{\sqrt{\lambda}}\Big).\end{align*}
The order $\sqrt{\lambda}^{0}$ piece can be made to match the one-loop
worldsheet result by setting \cite{McLoughlin:2008he} \begin{equation}
c=-\frac{\log2}{2\pi}.\label{eq:value-c-from-MRT}\end{equation}

The other way to resolve this is to modify the mode sum used. The
simplest object from the worldsheet perspective, and that used by
\cite{McLoughlin:2008ms,Alday:2008ut,Krishnan:2008zs,McLoughlin:2008he},
is \[
\delta E_{\mathrm{old}}\equiv\lim_{N\to\infty}\frac{1}{2}\sum_{n=-N}^{N}\omega_{n}\]
stopping at the same mode number $N$ for all modes. However, a different
cutoff is more natural when computing these modes using the algebraic
curve, namely to stop at a fixed radius $\left|x\right|$ in the spectral
plane. This new prescription was shown by \cite{Gromov:2008fy} to
change the result of \cite{McLoughlin:2008ms,Alday:2008ut,Krishnan:2008zs}
to\[
\delta E_{\mathrm{new}}=-3\frac{\log2}{2\pi}\log S\]
thus matching the Bethe ansatz calculation with $c=0$.

The fact that these two summation prescriptions (or regularisation
schemes) give different results can perhaps be summarised by saying
that these schemes refer to different coupling constants related by%
\footnote{We thank a referee for pointing this out. As noted by \cite{McLoughlin:2008he},
it is not clear whether or how $\lambda$ should be simultaneously
changed at weak coupling. %
} \[
\frac{1}{\sqrt{\lambda}}\;\to\;\frac{1}{\sqrt{\lambda}}\pm\frac{1}{\lambda}\left(\frac{\log2}{2\pi}\right)\]
This is clearly equivalent to changing $c$ in the expansion of $h(\lambda)$.

However it is not \emph{a priori} obvious that changing the cutoff
prescription from old to new is \emph{always} equivalent to such a
change of $\lambda$, or of $c$. What we will show here is that this
is also the case for energy corrections to giant magnons. But there
are of course many other one-loop calculations possible, all of which
are potentially affected.

We note that this scheme-dependence is not inherently an AdS/CFT issue:
we could see the changes in $\delta E$ for these string solutions
in $AdS_{4}\times CP^{3}$ even if we were unaware of the correspondence.
We would then call these terms $\alpha'$ corrections, and would see
no reason to expect them to be scheme-dependent. In a separate issue,
the $AdS_{4}$ radius $R$ receives corrections starting at two loops
\cite{Bergman:2009zh}, see also comments in\cite{McLoughlin:2008he}.
Neither of these issues occur in $AdS_{5}\times S^{5}$.

For now however we focus on the technical issues of these prescriptions,
returning to the larger discussion in the conclusion (section \ref{sub:conclusions-infinite-J}).

\subsection{Heavy and light modes\label{sub:Heavy-and-light-modes}}

The reason these two cutoff prescriptions differ is the existence
of a distinction between heavy and light modes. One sketch of why
this exists is to note that instead of $AdS_{5}\times S^{5}$ with
both spaces of radius $R$, we now have $AdS_{4}$ of radius $R/2$,
while $CP^{3}$ contains sphere-like subspaces of radius $R/2$ (namely
$CP^{1}$) and $R$ ($RP^{3}$), among other things. We expect that
the modes exploring this $RP^{3}$ should be lighter than those exploring
the $CP^{1}$ and $AdS_{4}$ directions. And indeed this is the case,
as can be seen directly \cite{Abbott:2008qd,Bandres:2009kw} or by
studying the Penrose limit \cite{Nishioka:2008gz,Gaiotto:2008cg,Grignani:2008is}.
The fermionic modes similarly fall into heavy and light groups. 

In the algebraic curve, we study modes by adding new poles to a pair
of quasimomenta. The position of these poles in the spectral plane
is governed by $q_{i}(x_{n})-q_{j}(x_{n})=2\pi n$, where $n\in\mathbb{Z}$
is the mode number. In $AdS_{5}\times S_{5}$, the vacuum has $q_{i}(x)=\alpha x/(x^{2}-1)$
for all $i$, and so the poles are always at \[
x_{n}^{\mathrm{heavy}}=\frac{\alpha}{4\pi n}+\sqrt{1+\left(\frac{\alpha}{4\pi n}\right)^{2}}.\]
But in $AdS_{4}\times CP^{3}$, the vacuum has $q_{i}(x)=\alpha x/(x^{2}-1)$
for $i=1,2,3,4$, but $q_{5}(x)=0$. The light modes are those in
which one of the quasimomenta involved is $q_{5}$ (or $q_{6}=-q_{5}$);
the others are heavy. The positions of their poles are related by
\[
x_{2n}^{\mathrm{heavy}}=x_{n}^{\mathrm{light}}.\]
This is exactly true for the vacuum, but will be approximately true
for fluctuations about arbitrary solutions, when $n$ is very large.
Thus we see that cutting off the sum at fixed $\left|x\right|$ is
amounts to cutting it off at $N$ for heavy modes but $N/2$ for light
modes:\[
\delta E_{\mathrm{new}}\equiv\lim_{\epsilon\to0}\frac{1}{2}\sum_{ij}\sum_{\left|x_{n}^{ij}\right|>1+\epsilon}\omega_{n}^{ij}=\lim_{N\to\infty}\frac{1}{2}\left(\sum_{n=-N}^{N}\omega_{n}^{\mathrm{heavy}}+\sum_{n=-N/2}^{N/2}\omega_{n}^{\mathrm{light}}\right).\]
This is new sum proposed by \cite{Gromov:2008fy}.

An alternative sum was proposed by \cite{Bandres:2009kw}, which uses
the same cutoff but omits the odd-numbered heavy modes: it can be
obtained from the `new' sum by replacing $\omega_{2n}+\omega_{2n+1}\to2\omega_{2n}$
for the heavy modes. 

Note that choosing which sum to perform is independent of choosing
whether to work with the algebraic curve or the worldsheet action,
as was stressed by \cite{Bandres:2009kw}. We would like to have a
physical reason for choosing one or the other.

\subsection{Giant magnons\label{sub:Giant-magnons-intro}}

The variety of sphere-like subspaces mentioned above allows a variety
of giant magnon solutions. The one whose dispersion relation we wrote
above is the elementary dyonic giant magnon \cite{Abbott:2009um,Hollowood:2009sc},
which explores a subspace $CP^{2}$. When $Q=1$ this reduces to an
embedding of the Hofman--Maldacena solution \cite{Hofman:2006xt}
into $CP^{1}$ \cite{Gaiotto:2008cg}. 

The other kinds of magnons are now understood to be superpositions
of two elementary magnons \cite{Hollowood:2009sc}. One choice of
orientations leads to an embedding of Dorey's dyonic magnon \cite{Dorey:2006dq,Chen:2006gea}
into $RP^{3}$, while another choice leads to a solution in which
the angular momenta $\pm Q$ cancel, leading to a two-parameter one-charge
solution we will refer to as the big giant magnon \cite{Hollowood:2009tw,Kalousios:2009mp,Suzuki:2009sc,Hatsuda:2009pc}.
When $Q\ll\sqrt{\lambda}$, both of these solutions reduce to an embedding
of the simple Hofman--Maldacena magnon into $RP^{2}$.

We can identify exactly the same states in the algebraic curve \cite{Shenderovich:2008bs,Lukowski:2008eq,Abbott:2009um}.
This is a convenient formalism for studying their semiclassical quantisation
--- constructing modes in the worldsheet theory is much more difficult
than for $AdS$ spinning strings \cite{Papathanasiou:2007gd}. Expanding
the magnon dispersion relation \eqref{eq:disp-rel-ABJM} in $1/\sqrt{\lambda}$,
for $Q=1$, \begin{align}
E & =\sqrt{\frac{1}{4}+4h(\lambda)^{2}\sin^{2}\frac{p}{2}}\nonumber \\
 & =\sqrt{2\lambda}\sin\frac{p}{2}+2c\sin\frac{p}{2}+\mathcal{O}\Big(\frac{1}{\sqrt{\lambda}}\Big)\label{eq:E-expansion}\\
 & =E_{\mathrm{class}}+\delta E+\ldots\nonumber \end{align}
we see that the one-loop correction $\delta E$ will teach us about
$c$. This is one reason for studying the semiclassical quantisation
of giant magnons. 

The first paper to calculate such a correction was \cite{Shenderovich:2008bs},
finding that, for the big giant magnon,\[
\delta E=0\]
consistent with $c=0$ (and exactly as in $AdS_{5}\times S^{5}$).
Since this paper pre-dated \cite{Gromov:2008fy}'s new sum prescription,
there appeared to be some tension with the $AdS$-sector results above.
However we show, by reverse-engineering, that the sum used is in fact
the new sum, and also that the result is the same for the elementary
magnon. We then perform the old sum, and find that instead\[
\delta E_{\mathrm{old}}=-2\frac{\log2}{2\pi}\sin\frac{p}{2}\]
implying the same $c=-\log2/2\pi$ as was found by \cite{McLoughlin:2008he}.
The results for the dyonic giant magnon (see \eqref{eq:small-mag-DeltaE-old}
below) and for various two--elementary-magnon solutions (appendices
\ref{sec:Corrections-Big} and \ref{sec:Corrections-RP3}) also point
to the same values for $c$. 

Our results for this $CP^{3}$ sector are thus in all cases consistent
with those found for the $AdS$ spinning strings. This still leaves
the value of $c$ apparently prescription-dependent. We comment further
on this in the conclusions.

\subsection{Outline}

In section \ref{sec:Semiclassical-using-Curves} we set up the machinery
for quantum corrections using the algebraic curve, using the off-shell
technique, and including the various summation prescriptions. We use
this in section \ref{sec:Elementary-Maganon-infinite-J} to calculate
corrections for the elementary giant magnon, including one kind of
finite-$J$ correction, the F-terms. We summarise and discuss our
results, as well as future directions, in section \ref{sec:Conclusions}.

Appendix \ref{sec:The-Classical-Algebraic-Curve} has some formulae
about the classical algebraic curve. Appendices \ref{sec:Corrections-Big}
and \ref{sec:Corrections-RP3} treat the `big' and $RP^{3}$ giant
magnons. Appendix \ref{sub:Vacuum-BL} is a note on conventions, and
appendix \ref{sec:Momentum-Conservation-etc} a note about momentum
conservation and level matching.

\section{Semiclassical Corrections using the Algebraic Curve\label{sec:Semiclassical-using-Curves}}

The classical algebraic curve is described by ten quasimomenta $q_{i}(x)$,
which are functions of the complex spectral parameter. We will be
concerned with a small perturbation of these to\[
q_{i}(x)+\delta q_{i}(x).\]
The perturbation $\delta q_{i}(x)$ inherits many properties from
the classical curve, in particular that only five of the ten sheets
are independent:\begin{equation}
\left(\vphantom{\frac{a}{a}}\delta q_{10},\delta q_{9},\delta q_{8},\delta q_{7},\delta q_{6}\right)=-\left(\vphantom{\frac{a}{a}}\delta q_{1},\delta q_{2},\delta q_{3},\delta q_{4},\delta q_{5}\right).\label{eq:q_eleven-minus-i-equals-q_i}\end{equation}
We summarise the other properties of the classical curve in appendix
\ref{sec:The-Classical-Algebraic-Curve}. Semiclassical methods presented
here originate in \cite{Gromov:2007aq,Beisert:2003xu,Beisert:2005di,Gromov:2008ec}.

\subsection{Perturbing the quasimomenta}

Fluctuations about the classical solution take the form of extra poles,
always appearing on a pair of sheets $(i,j)$. Those involving only
sheets 1,2 (or 9,10) represent bosonic fluctuations in $AdS_{4}$,
those involving only sheets 3,4,5 (or 6,7,8) bosonic fluctuations
in $CP^{3}$, and those which connect $AdS$ sheets to $CP$ sheets
fermionic fluctuations. We divide these fluctuations into light modes,
in which one of the sheets is 5 or 6, and heavy modes, the rest. Clearly
all the $AdS$ modes are heavy, but the $CP$ modes and fermions are
mixed. We refer to $(i,j)$ as the polarisation of the fluctuation;
here is a table of its possible values:%
\footnote{Note that we label all of these $(i,j)$ with $i<j$. Thanks to \eqref{eq:q_eleven-minus-i-equals-q_i}
the mode $(i,j)$ is equivalent to $(11-j,11-i)$, so we may also
always choose $i\leq5$. It will sometimes be convenient to define
$N_{11-j,11-i}=N_{ij}$, but $\sum_{ij}$ is always over the pairs
in this table.%
}

\begin{equation}\label{eq-tab:heavy-and-light-ads-and-cp}\begin{tabular}{c|ccc}
 & $AdS$ & Fermions & $CP$\tabularnewline
\hline
Heavy & (1,10) (2,9) (1,9) & (1,7) (1,8) (2,7) (2,8) & (3,7)\tabularnewline
Light &  & (1,5) (1,6) (2,5) (2,6) & (3,5) (3,6) (4,5) (4,6)\tabularnewline
\end{tabular}\end{equation}

The positions of these new poles, $x_{n}^{ij}$, satisfy\begin{equation}
q_{i}(x_{n}^{ij})-q_{j}(x_{n}^{ij})=2\pi n.\label{eq:pole-position}\end{equation}
Here $n$ is the mode number of the excitation, $N_{n}^{ij}$ is the
number of such excitations we turn on, and $N_{ij}=\sum_{n}N_{n}^{ij}$.
The level matching condition reads \begin{equation}
\sum_{n=-\infty}^{\infty}\sum_{ij}n\, N_{n}^{ij}=0.\label{eq:level-matching}\end{equation}

The residue at the new pole is fixed (in terms of its position) by
\begin{equation}
\delta q_{i}(x)=\frac{k_{ij}N_{n}^{ij}\alpha(x_{n}^{ij})}{x-x_{n}^{ij}}+\mathcal{O}(x-x_{n}^{ij})^{0}\label{eq:fluctuation-pole}\end{equation}
where \begin{equation}
\alpha(y)=\frac{1}{2g}\,\frac{y^{2}}{y^{2}-1}\label{eq:residue-alpha}\end{equation}
and the coefficients $k_{ij}$ are $\pm1$ or $\pm2$, to be read
off from \eqref{eq:delta-q-asympt-Nij} below. 

In addition to these new poles, $\delta q$ may also change the residues
at $x=\pm1$ provided these remain synchronised, and may shift endpoints
of the giant magnon's log cut (which is defined in \eqref{eq:defn-Gmag}
below). We will write these terms as \[
\delta q_{i}=\sum_{\pm}\frac{a_{\pm}}{x\pm1},\qquad i=1,2,3,4\]
and\begin{equation}
\delta q_{i}=\sum_{\pm}\frac{A^{\pm}}{x-X^{\pm}}\equiv M(x)\label{eq:defn-M-func}\end{equation}
which comes from $M(x)=-iA^{+}\frac{\partial}{\partial X^{+}}G_{\mathrm{mag}}(x)+iA^{-}\frac{\partial}{\partial X^{-}}G_{\mathrm{mag}}(x)$,
and so is added wherever the classical $q_{i}(x)$ contains the log
cut resolvent $G_{\mathrm{mag}}(x)$. 

The perturbation must also obey the inversion symmetries:\begin{align}
\delta q_{1}(\tfrac{1}{x}) & =-\delta q_{2}(x)\nonumber \\
\delta q_{3}(\tfrac{1}{x}) & =-\delta q_{4}(x)\label{eq:dq-inversion-symm}\\
\delta q_{5}(\tfrac{1}{x}) & =\delta q_{5}(x).\nonumber \end{align}
Note that the second of these imposes that there is no change in the
the total momentum $p$. Any momentum $\delta p$ carried by the fluctuation
must be cancelled by the change in the magnon's momentum, encoded
in $A^{\pm}$.

The change in the asymptotic charges is as follows:%
\footnote{Strictly speaking, for the sum on $j$ to be defined, we must interpret
$N_{ij}$ for $i>j$. For definiteness we adopt, here and in \eqref{eq:fluctuation-pole},
the convention that both $N_{ij}$ and $k_{ij}$ are symmetric. Our
signs for the asymptotic $\delta q$ match those of \cite{Shenderovich:2008bs};
in \cite{Bandres:2009kw} the signs of the fermions in $\delta q_{5}$
are reversed to $-N_{15}+N_{16}-N_{25}+N_{26}$.%
}\begin{align}
\delta q_{i} & \to\frac{1}{2gx}\sum_{j}k_{ij}N_{ij}+\begin{cases}
\frac{1}{2gx}\delta\Delta, & i=1\mbox{ or }2\\
0, & \mbox{otherwise}\end{cases}\quad\mbox{as }x\to\infty\displaybreak[0]\label{eq:delta-q-asympt-Nij}\\
 & =\frac{1}{2gx}\left(\begin{array}{ccc}
\delta\Delta+N_{19}+2N_{1\,10} & +N_{15}+N_{16}+N_{17}+N_{18}\\
\delta\Delta+2N_{29}+N_{19} & +N_{25}+N_{26}+N_{27}+N_{28}\\
 & -N_{18}-N_{28} & -N_{35}-N_{36}-N_{37}\\
 & -N_{17}-N_{27} & -N_{45}-N_{46}-N_{37}\\
 & +N_{15}-N_{16}+N_{25}-N_{26} & +N_{35}-N_{36}+N_{45}-N_{46}\end{array}\right).\nonumber \end{align}
For our purposes the energy shift $\delta\Delta$ is the output of
this calculation in which we constructed $\delta q_{i}(x)$. We define
the frequency $\Omega_{ij}(x_{n}^{ij})=\omega_{n}^{ij}$ of the $(i,j)$
mode to be $\delta\Delta$ when only that one fluctuation is turned
on, i.e. $N_{n}^{ij}=1$, others zero. This would however break \eqref{eq:level-matching},
so it is better to write\begin{equation}
\delta\Delta=\sum_{ij,n}N_{n}^{ij}\Omega_{ij}(x_{n}^{ij}).\label{eq:delta-Delta-and-sum-Omega}\end{equation}

\subsection{Off-shell method\label{sub:Off-shell-method}}

An efficient technique for calculating frequencies was invented by
\cite{Gromov:2008ec}, and adapted most explicitly to the $AdS_{4}\times CP^{3}$
case by \cite{Bandres:2009kw}. The idea is to temporarily ignore
condition \eqref{eq:pole-position} for the position of the new pole,
and place it at an arbitrary position $y$. The result is called an
off-shell perturbation, and we are interested in its frequency $\Omega_{ij}(y)$.
Having found a perturbation $\delta q$ for some polarisation $(i,j)$,
obeying all the conditions except \eqref{eq:pole-position}, we can
then use the inversion relations (as well as simply addition) to generate
such perturbations for other polarisations, along with their associated
frequencies. 

In fact knowing just two polarisations $(1,5)$ and $(4,5)$ is enough
to generate all the rest \cite{Bandres:2009kw}. First we use the
inversion conditions to obtain%
\footnote{This differs from \cite{Bandres:2009kw}'s equation (31b) thanks to
our conventions in \eqref{eq:delta-Delta-and-sum-Omega} above, see
appendix \ref{sub:Vacuum-BL}.%
} \begin{align}
\Omega_{25}(y) & =\Omega_{15}(0)-\Omega_{15}(\tfrac{1}{y})\label{eq:generate-25-35}\\
\Omega_{35}(y) & =\Omega_{45}(0)-\Omega_{45}(\tfrac{1}{y}).\nonumber \end{align}
(Here to construct $\delta^{25}q$ with a pole at $\left|y\right|>1$,
we must start with $\delta^{15}q$ with a pole inside the unit circle.)
The remaining light modes are simply given by $\delta q_{6}=-\delta q_{5}$,
thus\[
\Omega_{i6}(y)=\Omega_{i5}(y).\]
The heavy modes' frequencies are each the sum of two light modes',
since if we add $\delta^{i5}q+\delta^{5j}q$ (that is we switch on
$N_{i5}=1$ and $N_{5j}=N_{11-j,6}=1$) then the poles on sheets 5
and 6 will cancel. We obtain:\begin{align}
\Omega_{29}(y) & =2\Omega_{25}(y) & \Omega_{27} & =\Omega_{25}+\Omega_{45} & \Omega_{37} & =\Omega_{35}+\Omega_{45}.\nonumber \\
\Omega_{1\,10} & =2\Omega_{15} & \Omega_{17} & =\Omega_{15}+\Omega_{45}\label{eq:generate-lots-inv}\\
\Omega_{19} & =\Omega_{15}+\Omega_{25} & \Omega_{28} & =\Omega_{25}+\Omega_{35}\nonumber \\
 &  & \Omega_{18} & =\Omega_{15}+\Omega_{35}\nonumber \end{align}

Finally, we must then find the allowed poles $y=x_{n}^{ij}$ for each
polarisation. Evaluating the frequencies at these points gives us
the `on-shell' frequencies \begin{equation}
\omega_{n}^{ij}=\Omega_{ij}(x_{n}^{ij}).\label{eq:defn-on-shell-omega}\end{equation}
Note that for heavy modes, while the off-shell frequencies are always
the sum of two of those for light modes, the \emph{on-shell} frequencies
are not. We only expect the frequency to decompose $w_{m+n}^{ij}=w_{m}^{i5}+\omega_{n}^{5j}$
when the pole positions of the heavy and the two light modes happen
to agree: $x_{m+n}^{ij}=x_{m}^{i5}=x_{n}^{5j}$. This occurs for the
vacuum solution, see \eqref{eq:pole-position-vacuum} below, but not
for nontrivial classical solutions.

\subsection{Summing frequencies\label{sub:Definitions-of-Sums}}

The one-loop energy correction is given by \begin{align*}
\delta E & =\frac{1}{2}\sum_{ij,n}(-1)^{F_{ij}}\omega_{n}^{ij}, & F_{ij} & =\begin{cases}
0\\
1\end{cases}\mbox{for }(i,j)\;\begin{array}{l}
\mbox{bosonic}\\
\mbox{fermionic}.\end{array}\end{align*}
The way in which we deal with the infinite sum over $n$ is important,
and three different prescriptions have been given in the literature: 
\begin{enumerate}
\item The na\"{\i}ve sum cuts off at a fixed mode number $N$: \begin{align}
\delta E_{\mathrm{old}} & =\lim_{N\to\infty}\sum_{n=-N}^{N}\sum_{ij}(-1)^{F_{ij}}\frac{1}{2}\omega_{n}^{ij}\label{eq:sum-naiive}\\
 & =\lim_{N\to\infty}\frac{1}{2}\sum_{n=-N}^{N}\left(\vphantom{\frac{a}{a}}\omega_{n}^{\mathrm{heavy}}+\omega_{n}^{\mathrm{light}}\right).\nonumber \end{align}
This prescription makes no use of the distinction between heavy and
light modes, and is thus natural from the worldsheet perspective.
It was used by \cite{McLoughlin:2008ms,Alday:2008ut,Krishnan:2008zs}
for spinning string calculations. We have defined \cite{Gromov:2008fy,Bandres:2009kw}\begin{align}
\omega_{n}^{\mathrm{heavy}} & =w_{n}^{19}+w_{n}^{29}+w_{n}^{1\,10}+w_{n}^{37}-w_{n}^{17}-w_{n}^{18}-w_{n}^{27}-w_{n}^{28}\label{eq:defn-w-heavy-w-light}\\
\omega_{n}^{\mathrm{light}} & =w_{n}^{35}+w_{n}^{36}+w_{n}^{45}+w_{n}^{46}-w_{n}^{15}-w_{n}^{16}-w_{n}^{25}-w_{n}^{26}.\nonumber \end{align}

\item The sum proposed by Gromov and Mikhaylov \cite{Gromov:2008fy} is
this: \begin{align}
\delta E_{\mathrm{new}} & =\lim_{N\to\infty}\frac{1}{2}\sum_{m=-N}^{N}K_{m}\qquad\qquad\mbox{where }K_{m}=\begin{cases}
\omega_{m}^{\mathrm{heavy}}+\omega_{m/2}^{\mathrm{light}}, & m\mbox{ even}\\
\omega_{m}^{\mathrm{heavy}}, & m\mbox{ odd}\end{cases}\nonumber \\
 & =\lim_{N\to\infty}\frac{1}{2}\left(\sum_{n=-N}^{N}\omega_{n}^{\mathrm{heavy}}+\sum_{n=-N/2}^{N/2}\omega_{n}^{\mathrm{light}}\right).\label{eq:sum-new-GM}\end{align}
One justification for this change is that it amounts to including
all modes within some area of the spectral plane: at large $n$,\[
x_{n}^{ij}\approx\begin{cases}
1+\frac{\alpha}{4\pi n}+\mathcal{O}(\frac{\alpha}{n})^{2},\quad & (i,j)\mbox{ heavy}\\
1+\frac{\alpha}{8\pi n}+\mathcal{O}(\frac{\alpha}{n})^{2}, & (i,j)\mbox{ light}\end{cases}\]
so the last modes included in each sum, $x_{N}^{ij}$ (heavy) and
$x_{N/2}^{ij}$ (light) are at approximately the same position $x=1+\epsilon=1+\frac{\alpha}{4\pi N}$
in the spectral plane. In this sense it is natural from the algebraic
curve perspective. 
\item The sum proposed by Bandres and Lipstein \cite{Bandres:2009kw} is\begin{align}
\delta E_{\mathrm{new}'} & =\lim_{N\to\infty}\frac{1}{2}\sum_{m'=-N}^{N}\left(2\omega_{2m'}^{\mathrm{heavy}}+\omega_{m'}^{\mathrm{light}}\right)\label{eq:sum-new-BL}\\
 & =\lim_{N\to\infty}\frac{1}{2}\left(\sum_{\substack{n=-2N\\
n\:\mathrm{even}}
}^{2N}2\omega_{n}^{\mathrm{heavy}}+\sum_{n=-N}^{N}\omega_{n}^{\mathrm{light}}\right).\nonumber \end{align}
Unlike \cite{Gromov:2008fy}'s new sum above, this alternative new
sum has no odd-numbered heavy modes. In the continuum limit in which
$\delta E_{\mathrm{old}}=\int_{-\infty}^{\infty}dn\left(\omega_{n}^{\mathrm{heavy}}+\omega_{n}^{\mathrm{light}}\right)$,
both of the new prescriptions will agree: \begin{equation}
\delta E_{\mathrm{new}'}=\delta E_{\mathrm{new}}=\frac{1}{2}\int_{-\infty}^{\infty}dm\left(\omega_{m}^{\mathrm{heavy}}+\frac{1}{2}\omega_{m/2}^{\mathrm{light}}\right).\label{eq:sum-new-as-integral}\end{equation}
We discuss below another sense in which the two become equivalent,
at leading order \eqref{eq:deltaE-new2-BL}, although at subleading
order \eqref{eq:deltaE-subleading-BL} we can distinguish them. In
\eqref{eq:delta-E-F1-alternative-new-sum} we find a mismatch with
the L\"{u}scher F-term result of \cite{Bombardelli:2008qd}.
\end{enumerate}

\subsection[The vacuum]{Corrections for the vacuum\label{sub:Corrections-to-Vacuum}}

For the very simplest solution, we can evaluate these sums directly,
and always get zero. This solution is the BMN point particle, which
is the vacuum for giant magnons in the sense that it is dual to the
vacuum state of the spin chain. The classical curve is \cite{Gromov:2008bz}\begin{align*}
q_{1}(x) & =q_{2}(x)=q_{3}(x)=q_{4}(x)=\alpha\frac{x}{x^{2}-1}\\
q_{5}(x) & =0\end{align*}
where $\alpha=\Delta/2g$. The on-shell pole positions implied by
\eqref{eq:pole-position} are very simple, \begin{align}
x_{n}^{\mathrm{heavy}} & =\frac{\alpha}{4\pi n}\pm\sqrt{1+\left(\frac{\alpha}{4\pi n}\right)^{2}}\equiv V(n)\nonumber \\
x_{n}^{\mathrm{light}} & =V(2n)\label{eq:pole-position-vacuum}\end{align}
and we always choose the sign $\pm$ to maximise $\left|x_{n}\right|$.
Then have $x_{-n}=-x_{n}$. This fact is useful when constructing
the perturbation $\delta q_{i}$, as it allows one to use of a pair
of poles at $\pm y$, as was done by \cite{Gromov:2008bz,Bandres:2009kw}.
(See appendix \ref{sec:Momentum-Conservation-etc} for discussion.)
The first two off-shell frequencies are given by\begin{equation}
\Omega_{15}(y)=\Omega_{45}(y)=\frac{1}{y^{2}-1}.\label{eq:Omega-15-45-vac}\end{equation}
Using the results of section \ref{sub:Off-shell-method}, the others
are given (in our conventions) simply by \begin{equation}
\Omega_{ij}(y)=\begin{cases}
\frac{1}{y^{2}-1}, & (i,j)\mbox{ light}\\
\frac{2}{y^{2}-1}, & (i,j)\mbox{ heavy}.\end{cases}\label{eq:Omega-vac}\end{equation}
These lead to on-shell frequencies\begin{equation}
\omega_{n}^{ij}=\Omega_{ij}(x_{n}^{ij})=\begin{cases}
\sqrt{1+(\frac{4\pi}{\alpha})^{2}n^{2}}-1, & (i,j)\mbox{ heavy}\\
\sqrt{\frac{1}{4}+(\frac{4\pi}{\alpha})^{2}n^{2}}-\frac{1}{2}, & (i,j)\mbox{ light}.\end{cases}\label{eq:w_n-vac}\end{equation}

Similar frequencies can be found in the worldsheet theory. The precise
constant shifts ($-1$ and $-\frac{1}{2}$ here) of these are a matter
of convention in both the worldsheet and algebraic curve calculations,
see appendix \ref{sub:Vacuum-BL} for details. 

Since there are equally many bosonic and fermionic heavy modes, and
likewise light modes, we have the following cancellation at each $n$:\[
\omega_{n}^{\mathrm{heavy}}=\omega_{n}^{\mathrm{light}}=0.\]
Then all three of the above sums give zero:\[
\delta E_{\mathrm{old}}=\delta E_{\mathrm{new}}=\delta E_{\mathrm{new}'}=0.\]

\subsection{Some complex analysis\label{sub:Some-complex-analysis}}

To evaluate these sums in nontrivial cases, we can use the fact that
$\cot(z)$ has poles at $z=\pi n$ with residue 1 to write%
\footnote{All of our contours are $\circlearrowleft$.%
}\[
\delta E=\frac{1}{4i}\oint_{\mathbb{R}}dn\sum_{ij}(-1)^{F_{ij}}\cot(\pi n)\Omega_{ij}(x_{n}^{ij}).\]
We write this first as if there was no distinction between heavy and
light modes, as in \cite{SchaferNameki:2006gk,Gromov:2008ie}; we
will be more careful about exactly which sum prescription we are describing
afterwards.

For a given polarisation $(i,j)$, $n$ and $x$ are related by \eqref{eq:pole-position},
so we can write\begin{equation}
dn=\frac{q'_{i}(x)-q'_{j}(x)}{2\pi}dx.\label{eq:dn-to-dx}\end{equation}
The contour in $x$ should enclose all poles $x=x_{n}^{ij}$, which
are along the real line at $\left|x\right|>1$:\begin{equation}
\delta E=\frac{1}{4i}\oint_{\mathbb{R}(\left|x\right|>1)}dx\;\sum_{ij}(-1)^{F_{ij}}\:\frac{q'_{i}(x)-q'_{j}(x)}{2\pi}\cot\Big(\frac{q_{i}(x)-q_{j}(x)}{2}\Big)\;\Omega_{ij}(x).\label{eq:delta-E-first-x-integral}\end{equation}

\begin{figure}
\begin{centering}
\includegraphics[width=9.5cm]{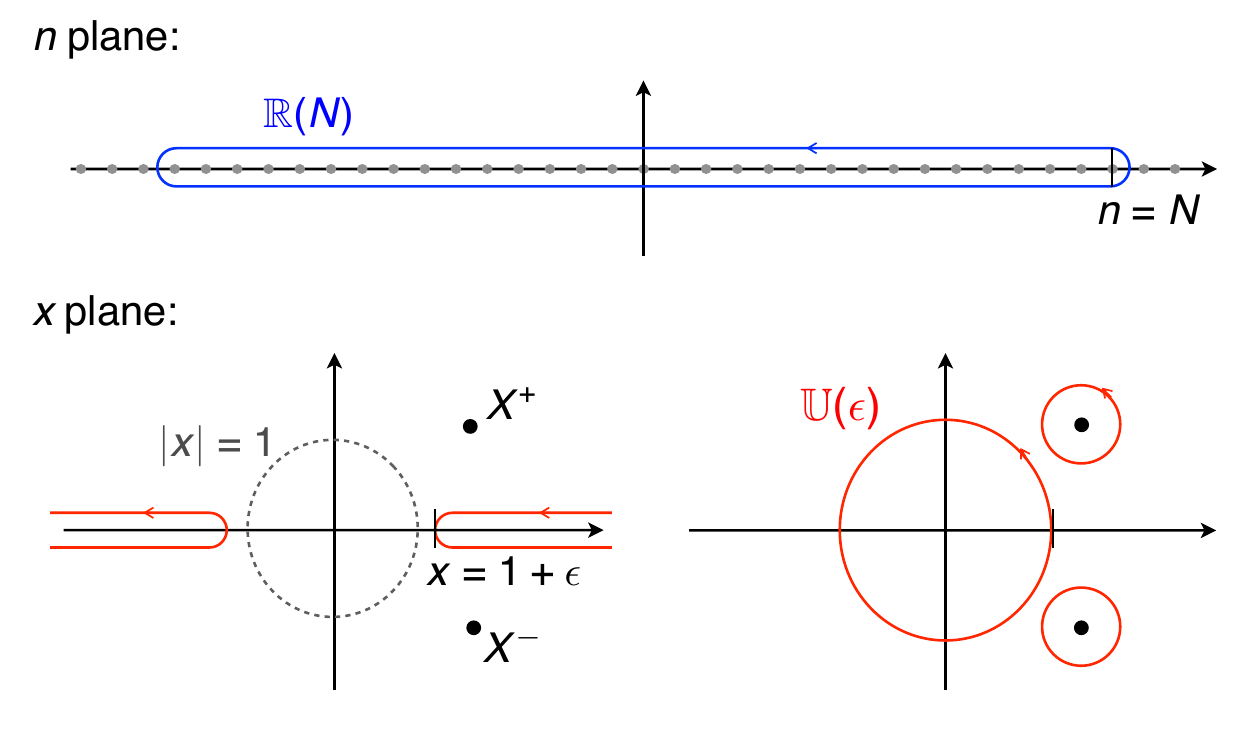}
\par\end{centering}

\caption{Integration contours in the complex $n$ and $x$ planes, showing
finite cutoffs $\left|n\right|\leq N$ and $\left|x\right|>1+\epsilon$.
(We do not attempt show the distinction between heavy and light modes
for the old and new old sums.) The first contour in the $x$ plane
is unwrapped to give the second, containing $\mathbb{U}(\epsilon)$,
after reversing its orientation. \label{fig:Integration-contours} }

\end{figure}

Next, deform the contour to one around the unit circle, in fact $-\mathbb{U}$
taking the orientation into account. (We draw the various contours
in figure \ref{fig:Integration-contours}.) There should be another
component around the branch points at $X^{\pm}$, but this is subleading,
and so we ignore it in this paper. Now write $\mathbb{U}=\mathbb{U}_{+}+\mathbb{U}_{-}$
for the parts of the unit circle above and below the real line. On
this circle $q_{i}-q_{j}$ is large, and so we can approximate\begin{align}
\cot\Big(\frac{q_{i}-q_{j}}{2}\Big) & =\pm i\left(1+2e^{\mp i(q_{i}-q_{j})}+2e^{\mp2i(q_{i}-q_{j})}+\ldots\right).\label{eq:cot-qq-expansion}\end{align}
We keep only the first term for now (returning to subsequent terms
in the next section):\[
\delta E\approx-\frac{1}{8\pi i}\sum_{\pm}\pm i\int_{\mathbb{U}_{\pm}}dx\sum_{ij}(-1)^{F_{ij}}\left[\vphantom{\frac{a}{a}}q'_{i}(x)-q'_{j}(x)\right]\Omega_{ij}(x).\]

In order to distinguish the old and new sums, we must be careful about
their upper limits. Let us write $\mathbb{R}(N)$ for the contour
encircling the integers up to $\pm N$, and $\mathbb{U}(\epsilon)$
for a unit circle at radius $1+\epsilon$.
\begin{itemize}
\item The new sum \eqref{eq:sum-new-GM} turns out to be the simplest case.
Following the above steps, we write:%
\footnote{If we define $z=n$ for the heavy modes but $z=2n$ for the light
modes, then we can also re-write the integrals over $n$ as one integral
over $z$:\[
dE_{\mathrm{new}}=\lim_{N\to\infty}\frac{1}{4i}\oint_{\mathbb{R}(N)}dz\left[\cot(\pi z)\,\omega_{z}^{\mathrm{heavy}}+\frac{1}{2}\cot\negmedspace\left(\frac{\pi z}{2}\right)\omega_{z/2}^{\mathrm{light}}\right].\]
This is perhaps more natural from the form of the sum $\delta E_{\mathrm{new}}=\sum_{m}K_{m}$
in \eqref{eq:sum-new-GM}. We stress however that \eqref{eq:pole-position}
and \eqref{eq:dn-to-dx} contain $n$, not $z$.%
} \begin{align}
\delta E_{\mathrm{new}} & =\lim_{N\to\infty}\frac{1}{2}\left(\sum_{n=-N}^{N}\omega_{n}^{\mathrm{heavy}}+\sum_{n=-N/2}^{N/2}\omega_{n}^{\mathrm{light}}\right)\nonumber \\
 & =\lim_{N\to\infty}\frac{1}{4i}\left(\oint_{\mathbb{R}(N)}dn\,\cot(\pi n)\,\omega_{n}^{\mathrm{heavy}}+\oint_{\mathbb{R}(N/2)}dn\,\cot(\pi n)\,\omega_{n}^{\mathrm{light}}\right)\displaybreak[0]\nonumber \\
 & =\lim_{\epsilon\to0}-\frac{1}{4i}\oint_{\mathbb{U}(\epsilon)}dx\sum_{ij}(-1)^{F_{ij}}\frac{q'_{i}(x)-q'_{j}(x)}{2\pi}\cot\left(\frac{q_{i}(x)-q_{j}(x)}{2}\right)\Omega_{ij}(x)\nonumber \\
 & \approx-\frac{1}{8\pi i}\sum_{\pm}\pm i\int_{\mathbb{U}_{\pm}}dx\sum_{ij}(-1)^{F_{ij}}\left[\vphantom{\frac{a}{a}}q'_{i}(x)-q'_{j}(x)\right]\Omega_{ij}(x).\label{eq:deltaE-new-GM}\end{align}
Since we have the same contour $\mathbb{U}(\epsilon)$ for both heavy
and light modes, we can write them as one integral. The last line
is the leading term in the expansion \eqref{eq:cot-qq-expansion}.
\item For the alternative new sum \eqref{eq:sum-new-BL}, \begin{align}
\delta E_{\mathrm{new}'} & =\lim_{N\to\infty}\frac{1}{2}\Bigg(\sum_{\substack{n=-2N\\
n\:\mathrm{even}}
}^{2N}2\omega_{n}^{\mathrm{heavy}}+\sum_{n=-N}^{N}\omega_{n}^{\mathrm{light}}\Bigg)\nonumber \\
 & =\lim_{N\to\infty}\frac{1}{4i}\left(\oint_{\mathbb{R}(2N)}dn\,\cot\negmedspace\left(\frac{\pi n}{2}\right)\omega_{n}^{\mathrm{heavy}}+\oint_{\mathbb{R}(N)}dn\,\cot(\pi n)\,\omega_{n}^{\mathrm{light}}\right)\displaybreak[0]\nonumber \\
 & =\lim_{\epsilon\to0}-\frac{1}{4i}\oint_{\mathbb{U}(\epsilon)}dx\sum_{ij}(-1)^{F_{ij}}\frac{q'_{i}(x)-q'_{j}(x)}{2\pi}\begin{cases}
\cot(\frac{q_{i}(x)-q_{j}(x)}{4})\Omega_{ij}(x), & (i,j)\mbox{ heavy}\\
\cot(\frac{q_{i}(x)-q_{j}(x)}{2})\Omega_{ij}(x), & (i,j)\mbox{ light}\end{cases}\nonumber \\
 & \approx-\frac{1}{8\pi i}\sum_{\pm}\pm i\int_{\mathbb{U}_{\pm}}dx\sum_{ij}(-1)^{F_{ij}}\left[\vphantom{\frac{a}{a}}q'_{i}(x)-q'_{j}(x)\right]\Omega_{ij}(x).\label{eq:deltaE-new2-BL}\end{align}
Notice that the difference between this and the new sum (i.e. the
argument of the cotangent) disappears in the leading term of \eqref{eq:cot-qq-expansion}. 
\item Finally, for the old sum \eqref{eq:sum-naiive}, \begin{align}
\negthickspace\negthickspace\delta E_{\mathrm{old}} & =\lim_{N\to\infty}\frac{1}{2}\sum_{n=-N}^{N}\left(\omega_{n}^{\mathrm{heavy}}+\omega_{n}^{\mathrm{light}}\right)\nonumber \\
 & =\lim_{N\to\infty}\frac{1}{4i}\oint_{\mathbb{R}(N)}dn\cot(\pi n)\left(\omega_{n}^{\mathrm{heavy}}+\omega_{n}^{\mathrm{light}}\right)\displaybreak[0]\nonumber \\
 & =\lim_{\epsilon\to0}\frac{-1}{4i}\Bigg\lbrace\oint_{\mathbb{U}(2\epsilon)}\negthickspace\negthickspace\negthickspace dx\negthickspace\sum_{\substack{ij\\
\mathrm{heavy}}
}+\oint_{\mathbb{U}(\epsilon)}\negthickspace\negthickspace\negthickspace dx\negthickspace\sum_{\substack{ij\\
\mathrm{light}}
}\Bigg\rbrace(-1)^{F_{ij}}\frac{q'_{i}(x)-q'_{j}(x)}{2\pi}\cot\left(\frac{q_{i}(x)-q_{j}(x)}{2}\right)\Omega_{ij}(x)\nonumber \\
 & \approx\lim_{\epsilon\to0}\left[\vphantom{\frac{a}{a}}L^{\mathrm{heavy}}(2\epsilon)+L^{\mathrm{light}}(\epsilon)\right]\label{eq:deltaE-old}\end{align}
where in the last line we write the leading term of \eqref{eq:cot-qq-expansion}
in terms of the integral\begin{flalign}
L^{\mathrm{light}}(\epsilon) & =-\frac{1}{8\pi i}\sum_{\pm}\pm i\int_{\mathbb{U}_{\pm}(\epsilon)}dx\sum_{\substack{ij\\
\mathrm{light}}
}(-1)^{F_{ij}}\left[\vphantom{\frac{a}{a}}q'_{i}(x)-q'_{j}(x)\right]\Omega_{ij}(x).\label{eq:defn-L-integral}\end{flalign}
We will explicitly perform this integral along contours of two different
radii, $1+\epsilon$ and $1+2\epsilon$, and add them before taking
the limit $\epsilon\to0$. 
\end{itemize}

\subsection{Subleading terms\label{sub:Subleading-terms}}

In \eqref{eq:deltaE-new-GM}, \eqref{eq:deltaE-new2-BL} and \eqref{eq:deltaE-old}
above, we kept only the first term in the expansion \eqref{eq:cot-qq-expansion}.
We now consider the next term, which we call $\delta E^{1}$.
\begin{itemize}
\item For the new sum, we have (integrating by parts) \begin{align}
\delta E_{\mathrm{new}}^{1} & =-\frac{1}{8\pi i}\sum_{\pm}\int_{\mathbb{U}_{\pm}}dx\sum_{ij}(-1)^{F_{ij}}\partial_{x}\left[q_{i}(x)-q_{j}(x)\right]\left(\pm i2e^{\mp i(q_{i}-q_{j})}\right)\Omega_{ij}(x)\nonumber \\
 & =-\frac{1}{4\pi i}\sum_{\pm}\int_{\mathbb{U}_{\pm}}dx\sum_{ij}(-1)^{F_{ij}}e^{\mp i(q_{i}-q_{j})}\partial_{x}\Omega_{ij}(x).\label{eq:deltaE-subleading-new}\end{align}

\item For the alternative new sum, the only change is in the exponent for
the heavy modes: \begin{align}
\negthickspace\negthickspace\negthickspace\negthickspace\delta E_{\mathrm{new}'}^{1} & =-\frac{1}{8\pi i}\oint_{\mathbb{U}}dx\sum_{ij}(-1)^{F_{ij}}\partial_{x}\left[q_{i}(x)-q_{j}(x)\right]\begin{cases}
\left(\pm i2e^{\mp i\frac{q_{i}-q_{j}}{2}}\right)\Omega_{ij}(x), & (i,j)\mbox{ heavy}\\
\left(\pm i2e^{\mp i(q_{i}-q_{j})}\right)\Omega_{ij}(x), & (i,j)\mbox{ light}\end{cases}\nonumber \\
 & \negthickspace\negthickspace\negthickspace\negthickspace=-\frac{1}{4\pi i}\sum_{\pm}\int_{\mathbb{U}_{\pm}}dx\Bigg[\sum_{\substack{ij\\
\mathrm{heavy}}
}(-1)^{F_{ij}}e^{\mp i\frac{q_{i}-q_{j}}{2}}2\partial_{x}\Omega_{ij}(x)+\sum_{\substack{ij\\
\mathrm{light}}
}(-1)^{F_{ij}}e^{\mp i(q_{i}-q_{j})}\partial_{x}\Omega_{ij}(x)\Bigg].\label{eq:deltaE-subleading-BL}\end{align}

\item And finally, for the old sum, the only difference from the new sum
is in the contour $\mathbb{U}(2\epsilon)$ for the heavy modes: \begin{align}
\delta E_{\mathrm{old}}^{1} & =-\frac{1}{4\pi i}\sum_{\pm}\Bigg\lbrace\int_{\mathbb{U}_{\pm}(2\epsilon)}\negthickspace\negthickspace dx\sum_{\substack{ij\\
\mathrm{heavy}}
}+\int_{\mathbb{U}_{\pm}(\epsilon)}\negthickspace\negthickspace dx\sum_{\substack{ij\\
\mathrm{light}}
}\Bigg\rbrace(-1)^{F_{ij}}e^{\mp i(q_{i}-q_{j})}\partial_{x}\Omega_{ij}(x).\label{eq:deltaE-subleading-old}\end{align}

\end{itemize}
We can continue with the higher terms in \eqref{eq:cot-qq-expansion},
and calling their total $\delta E^{F}$, write\begin{equation}
\delta E_{\mathrm{new}}^{F}=\sum_{m=1,2,3\ldots}\frac{-1}{4\pi i}\sum_{\pm}\int_{\mathbb{U}_{\pm}}dx\sum_{ij}(-1)^{F_{ij}}e^{\mp mi(q_{i}-q_{j})}\frac{1}{m}\partial_{x}\Omega_{ij}(x).\label{eq:dE-F-all-m}\end{equation}
Similar expressions can clearly be written down for the old and the
alternative new sums.

\section{Corrections for the Elementary Giant Magnon\label{sec:Elementary-Maganon-infinite-J}}

Here we study the solution constructed in the $\sigma$-model by \cite{Abbott:2009um,Hollowood:2009sc}
and in the algebraic curve by \cite{Gromov:2008bz,Shenderovich:2008bs}.
This is also known as the small or $CP^{2}$ giant magnon.

\subsection{Classical curve}

The magnon is described by the algebraic curve\begin{align}
q_{1}(x) & =\dfrac{\alpha x}{x^{2}-1}\nonumber \\
q_{2}(x) & =\dfrac{\alpha x}{x^{2}-1}\nonumber \\
q_{3}(x) & =\dfrac{\alpha x}{x^{2}-1}+G(0)-G(\tfrac{1}{x})-\frac{p}{2}\label{eq:Alg-curve-Small-mag}\\
q_{4}(x) & =\dfrac{\alpha x}{x^{2}-1}+G(x)-\frac{p}{2}\nonumber \\
q_{5}(x) & =G(x)-G(0)+G(\tfrac{1}{x})\nonumber \end{align}
where $p=-i\log(X^{+}/X^{-})$ and the resolvent is \cite{Minahan:2006bd,Vicedo:2007rp,Chen:2007vs}
\begin{equation}
G(x)=G_{\mathrm{mag}}(x)=-i\log\left(\frac{x-X^{+}}{x-X^{-}}\right).\label{eq:defn-Gmag}\end{equation}
Here we have included the twists in $q_{3}(x)$ and $q_{4}(x)$ as
used by \cite{Gromov:2008ie,Gromov:2007ky} which amount to orbifolding
the space by an angle $p$ so as to make the giant magnon a closed
string \cite{Astolfi:2007uz,Ideguchi:2004wm,Solovyov:2007pw}. Note
that these twists play no role in the leading corrections, but are
important in the subleading corrections.

The charges $E$ and $Q$ can be read off from the behaviour of $q(x)$
at infinity, see \eqref{eq:asympt-curve-charge} below, and are given
in terms of $X^{\pm}$ by\begin{align}
E=\Delta-\frac{J}{2} & =-ig\left(X^{+}-\frac{1}{X^{+}}-X^{-}+\frac{1}{X^{-}}\right)\nonumber \\
Q & =-i2g\left(X^{+}+\frac{1}{X^{+}}-X^{-}-\frac{1}{X^{-}}\right).\label{eq:Q-in-gXpXm}\end{align}
These can be combined to give the dispersion relation \eqref{eq:disp-rel-ABJM}.

\subsection{Off-shell frequencies\label{sub:finding-frequencies-small}}

For the $(1,5)$ polarisation we use the following ansatz, with $\alpha(y)$
and $M(x)$ defined in \eqref{eq:residue-alpha} and \eqref{eq:defn-M-func}
above:\begin{align}
\delta q_{1} & =\frac{\alpha(y)}{x-y}+\sum_{\pm}\frac{a_{\pm}}{x\pm1} & \negthickspace\negthickspace\negthickspace\negthickspace\negthickspace\negthickspace\negthickspace\negthickspace\negthickspace\negthickspace\delta q_{2} & =-\delta q_{1}(\tfrac{1}{x})\nonumber \\
\delta q_{4} & =\sum_{\pm}\frac{a_{\pm}}{x\pm1}+M(x) & \negthickspace\negthickspace\negthickspace\negthickspace\negthickspace\negthickspace\negthickspace\negthickspace\negthickspace\negthickspace\delta q_{3} & =-\delta q_{4}(\tfrac{1}{x})\label{eq:small-dq-ansatz}\\
\delta q_{5} & =\frac{\alpha(y)}{x-y}+\frac{\alpha(y)}{\frac{1}{x}-y}+\frac{\alpha(y)}{y}+M(x)-M(0)+M(\tfrac{1}{x}).\negthickspace\negthickspace\negthickspace\negthickspace\negthickspace\negthickspace\negthickspace\negthickspace\negthickspace\negthickspace & \negthickspace\negthickspace\negthickspace\negthickspace\negthickspace\negthickspace\negthickspace\negthickspace\negthickspace\negthickspace\nonumber \end{align}
This clearly has the correct new poles and satisfies the inversion
symmetries, and also has synchronised poles at $x=\pm1$. It remains
to impose the conditions at infinity, starting with the conditions
that $\delta q_{i}$ vanish there. The nontrivial ones are:%
\footnote{After satisfying these, we can write\begin{align*}
\delta q_{2}(x) & =\sum_{\pm}\frac{a_{\pm}}{x\pm1}+\frac{\alpha(y)/y^{2}}{x-\frac{1}{y}}\\
\delta q_{3}(x) & =\sum_{\pm}\frac{a_{\pm}}{x\pm1}+M(0)-M(\tfrac{1}{x}).\end{align*}
Here $\delta q_{2}$ has a new pole inside the unit circle, and $\delta q_{3}$
has the expected pattern of $M(x)$. %
} \begin{align*}
\delta q_{2}(\infty) & =\frac{\alpha(y)}{y}-a_{+}+a_{-}=0\\
\delta q_{3}(\infty) & =-M(0)-a_{+}+a_{-}=0\end{align*}
which, recalling that $M(x)=\sum_{\pm}A^{\pm}/(x-X^{\pm})$, imply
\begin{equation}
\frac{A^{+}}{X^{+}}+\frac{A^{-}}{X^{-}}=\frac{\alpha(y)}{y}.\label{eq:SM15-First-eq-for-A+A-}\end{equation}
Next, the $1/x$ behaviour gives the following equations (not all
independent):\begin{flalign*}
\delta q_{1}(x) & \sim\frac{1}{x}\left[\vphantom{\frac{a}{b}}a_{+}+a_{-}+\alpha(y)\right]\;=\frac{\delta\Delta+1}{2gx}\\
\delta q_{2}(x) & \sim\frac{1}{x}\left[a_{+}+a_{-}+\frac{\alpha(y)}{y^{2}}\right]\;=\frac{\delta\Delta}{2gx}\displaybreak[0]\\
\delta q_{3}(x) & \sim\frac{1}{x}\left[\vphantom{\frac{a}{b}}a_{+}+a_{-}+A^{+}+A^{-}\right]\;=0\displaybreak[0]\\
\delta q_{4}(x) & \sim\frac{1}{x}\left[a_{+}+a_{-}+\frac{A^{+}}{X^{+2}}+\frac{A^{-}}{X^{-2}}\right]\;=0\\
\delta q_{5}(x) & \sim\frac{1}{x}\left[-\alpha(y)+\frac{\alpha(y)}{y^{2}}+A^{+}+A^{-}-\frac{A^{+}}{X^{+2}}-\frac{A^{-}}{X^{-2}}\right]\;=-\frac{1}{2gx}.\end{flalign*}
Using the $\delta q_{2}$ and $\delta q_{4}$ equations we can write
$\delta\Delta$ in terms of $y$, $X^{\pm}$ and $A^{\pm}$:\begin{equation}
\frac{\delta\Delta}{2g}=\frac{\alpha(y)}{y^{2}}-\left(\frac{A^{+}}{X^{+2}}+\frac{A^{-}}{X^{-2}}\right).\label{eq:SM15-Energy-shift-def}\end{equation}
The $\delta q_{5}$ equation gives\begin{equation}
A_{+}+A_{-}=\frac{A^{+}}{X^{+2}}+\frac{A^{-}}{X^{-2}}\label{eq:SM15-Second-eq-for-A+A-}\end{equation}
which we can use with \eqref{eq:SM15-First-eq-for-A+A-} to find $A^{\pm}$:
first write \[
A^{+}=\frac{\alpha(y)}{y}X^{+}-\frac{A^{-}X^{+}}{X^{-}}.\]
and then plugging this into \prettyref{eq:SM15-Second-eq-for-A+A-}
we find\begin{align}
A^{-} & =-\frac{\alpha(y)}{y}\frac{\left(X^{+2}-1\right)X^{-2}}{\left(X^{-}-X^{+}\right)\left(X^{-}X^{+}+1\right)},\nonumber \\
A^{+} & =\frac{\alpha(y)}{y}X^{+}+\frac{\alpha(y)}{y}\frac{\left(X^{+2}-1\right)X^{-}X^{+}}{\left(X^{-}-X^{+}\right)\left(X^{-}X^{+}+1\right)}.\label{eq:SM15-Result-for-A+A-}\end{align}
We can now finally write the frequency $\Omega_{15}(y)$ from \prettyref{eq:SM15-Energy-shift-def}:\begin{flalign*}
\Omega_{15}(y)=\delta\Delta & =\frac{2g\alpha(y)}{y^{2}}\left[1+y\left(\frac{\left(X^{+2}-1\right)}{\left(X^{-}-X^{+}\right)\left(X^{-}X^{+}+1\right)}\left(1-\frac{X^{-}}{X^{+}}\right)-\frac{1}{X^{+}}\right)\right]\\
 & =\frac{1}{y^{2}-1}\left(1-y\frac{X^{+}+X^{-}}{X^{+}X^{-}+1}\right).\end{flalign*}

For the $(4,5)$ polarisation, we use a similar ansatz \begin{align*}
\delta q_{1} & =\sum_{\pm}\frac{a_{\pm}}{x\pm1} & \negthickspace\negthickspace\negthickspace\negthickspace\negthickspace\negthickspace\negthickspace\negthickspace\negthickspace\negthickspace\delta q_{2} & =-\delta q_{1}(\tfrac{1}{x})\\
\delta q_{4} & =-\frac{\alpha(x)}{x-y}+\sum_{\pm}\frac{a_{\pm}}{x\pm1}+M(x) & \negthickspace\negthickspace\negthickspace\negthickspace\negthickspace\negthickspace\negthickspace\negthickspace\negthickspace\negthickspace\delta q_{3} & =-\delta q_{4}(\tfrac{1}{x})\\
\delta q_{5} & =\frac{\alpha(y)}{x-y}+\frac{\alpha(y)}{\frac{1}{x}-y}+\frac{\alpha(y)}{y}+M(x)-M(0)+M(\tfrac{1}{x})\negthickspace\negthickspace\negthickspace\negthickspace\negthickspace\negthickspace\negthickspace\negthickspace\negthickspace\negthickspace & \negthickspace\negthickspace\negthickspace\negthickspace\negthickspace\negthickspace\negthickspace\negthickspace\negthickspace\negthickspace\end{align*}
and a similar computation leads to the same frequency:\[
\Omega_{45}(y)=\frac{1}{y^{2}-1}\left(1-y\frac{X^{+}+X^{-}}{X^{+}X^{-}+1}\right).\]

Constructing all the other frequencies using the formulae of section
\ref{sub:Off-shell-method}, we find simply:\begin{equation}
\Omega_{ij}(y)=\begin{cases}
\Omega_{45}(y), & (i,j)\mbox{ light}\\
2\Omega_{45}(y), & (i,j)\mbox{ heavy}.\end{cases}\label{eq:Omega-small}\end{equation}
This differs from the $AdS_{5}\times S^{5}$ case only by the factor
of 2 in the heavy modes, and agrees with the big giant magnon calculation
of \cite{Shenderovich:2008bs}.

\subsection{Leading energy corrections\label{sub:Leading-corrections-elementary-magnon}}

Here we calculate the integrals described in section \ref{sub:Some-complex-analysis}.
We begin by noting the following identity, which follows simply from
the list of possible polarisations \eqref{eq-tab:heavy-and-light-ads-and-cp}:\begin{flalign}
\sum_{\substack{ij\\
\mathrm{heavy}}
}(-1)^{F_{ij}}\left[\vphantom{\frac{a}{a}}q'_{i}-q'_{j}\right] & \;=\; q'_{1}+q'_{2}-q'_{3}-q'_{4}\;=\;-\frac{1}{2}\sum_{\substack{ij\\
\mathrm{light}}
}(-1)^{F_{ij}}\left[\vphantom{\frac{a}{a}}q'_{i}-q'_{j}\right].\label{eq:sum-qq-1234-identity}\end{flalign}
Using this result, the new sums are trivial, because thanks to the
factor 2 in \eqref{eq:Omega-small}, the integrands in both \eqref{eq:deltaE-new-GM}
and \eqref{eq:deltaE-new2-BL} vanish: \[
\sum_{ij}(-1)^{F_{ij}}\left[\vphantom{\frac{a}{a}}q'_{i}(x)-q'_{j}(x)\right]\Omega_{ij}(x)=0.\]
So we have, at leading order, \begin{equation}
\delta E_{\mathrm{new}}=\delta E_{\mathrm{new}'}=0.\label{eq:small-mag-DeltaE-new}\end{equation}

We wrote the old sum in \eqref{eq:deltaE-old} in terms of two integrals
$L^{\mathrm{w}}(\epsilon)$. Using identity \eqref{eq:sum-qq-1234-identity}
and $\Omega_{ij}(y)$ from \eqref{eq:Omega-small}, we see that \[
L^{\mathrm{heavy}}(\epsilon)=-L^{\mathrm{light}}(\epsilon).\]
Using the explicit classical curve \eqref{eq:Alg-curve-Small-mag},
we also have \[
q'_{1}(x)+q'_{2}(x)-q'_{3}(x)-q'_{4}(x)=i\left(\frac{1}{x-X^{+}}-\frac{1}{x-X^{-}}\right)-i\left(\frac{1}{xX^{+}-1}-\frac{1}{xX^{-}-1}\right).\]
We can now evaluate the integral explicitly, keeping $\epsilon$ finite:
parametrise $x=(1+\epsilon)e^{i\varphi}$, where $\varphi\in[0,\pi]$
in $\mathbb{U}_{+}(\epsilon)$ and $\varphi\in[\pi,2\pi]$ in $\mathbb{U}_{-}(\epsilon)$.
Then \begin{align*}
L^{\mathrm{light}}(\epsilon) & =\frac{-1}{8\pi i}\sum_{\pm}\pm i\int_{\mathbb{U}_{\pm}(\epsilon)}dx\sum_{\substack{ij\\
\mathrm{light}}
}(-1)^{F_{ij}}\left[\vphantom{\frac{a}{a}}q'_{i}(x)-q'_{j}(x)\right]\Omega_{ij}(x)\displaybreak[0]\\
 & =\frac{1}{\pi i(1+X^{+}X^{-})}\left[\vphantom{\frac{A}{B}}2(X^{-}-X^{+})\arctan(1+\epsilon)+2X^{+}X^{-}\arctan(X^{-}\left(1+\epsilon\right))\right.\\
 & \qquad+(X^{-}-X^{+})\log\left(\frac{-(2+\epsilon)}{\epsilon}\right)+X^{+}X^{-}\log\left(\frac{1-X^{+}(1+\epsilon)}{1+X^{+}(1+\epsilon)}\right)\\
 & \qquad\left.+\log\left(\frac{\left(1+\epsilon-X^{-}\right)}{\left(1+\epsilon+X^{-}\right)}\frac{\left(1+\epsilon+X^{+}\right)}{\left(1+\epsilon-X^{+}\right)}\right)\right].\end{align*}
Only one term diverges as $\epsilon\to0$. It is this divergent term
which makes the limit in \eqref{eq:deltaE-old} nontrivial, and which
leads to the following result for the leading term in $\delta E$:\begin{align}
\delta E_{\mathrm{old}} & =\lim_{\epsilon\to0}\left[\vphantom{\frac{a}{a}}L^{\mathrm{heavy}}(2\epsilon)+L^{\mathrm{light}}(\epsilon)\right]\nonumber \\
 & =\lim_{\epsilon\to0}\left[\vphantom{\frac{a}{a}}-L^{\mathrm{light}}(2\epsilon)+L^{\mathrm{light}}(\epsilon)\right]\nonumber \\
 & =\frac{i\log2}{\pi}\frac{X^{+}-X^{-}}{1+X^{+}X^{-}}.\label{eq:small-mag-DeltaE-old}\end{align}
In the non-dyonic case ($Q=1\ll\sqrt{\lambda}$ thus $X^{\pm}=e^{\pm ip/2}$)
this becomes \begin{equation}
\delta E_{\mathrm{old}}=-\frac{\log2}{\pi}\sin\frac{p}{2}.\label{eq:small-DeltaE-old-nondyonic}\end{equation}
Comparing to the expansion \eqref{eq:E-expansion} of the dispersion
relation, we recover \eqref{eq:value-c-from-MRT}: \[
c=-\frac{\log2}{2\pi}.\]

In summary, the situation for these leading corrections for the giant
magnon is exactly the same as for the leading corrections for spinning
strings in $AdS$. Either we use the old sum prescription and $c\neq0$,
or we use either of the new sums and $c=0$.

We repeat this analysis for the `big' and $RP^{3}$ giant magnons
in appendices \ref{sec:Corrections-Big} and \ref{sec:Corrections-RP3},
reaching the same conclusion in each case.

\subsubsection*{Dyonic case}

We could write the above result \eqref{eq:small-DeltaE-old-nondyonic}
as\begin{equation}
\delta E_{\mathrm{old}}-\delta E_{\mathrm{new}}=c\,\frac{1}{2}\frac{\partial}{\partial g}E\label{eq:old-new-derivative-g}\end{equation}
if we use the classical energy $E$ in terms of $g$ (i.e. in terms
of $\lambda$), recalling that $h(\lambda)=2g=\sqrt{\lambda/2}$ at
leading order. This is of course exactly the second term in the expansion
\eqref{eq:E-expansion}.

Now we observe that this is also true for the dyonic case, provided
we hold fixed $p$ and $Q$:\begin{align}
E & =\sqrt{\frac{Q^{2}}{4}+16g^{2}\sin^{2}\frac{p}{2}}\nonumber \\
\implies\quad\frac{\partial}{\partial g}E & =\frac{16g\sin^{2}\frac{p}{2}}{E}=-4i\frac{X^{+}-X^{-}}{1+X^{+}X^{-}}\label{eq:derivative-E-with-g-dyonic}\end{align}
correctly reproducing \eqref{eq:small-mag-DeltaE-old}. We have used
\eqref{eq:Q-in-gXpXm} to write this in $X^{\pm}$, but we stress
that the derivative is not holding $X^{\pm}$ fixed. 

We note that the dyonic giant magnon is the first example for which
the classical energy which one expands is not proportional to $h(\lambda)$.
This fact made the expansions for both the $AdS_{3}$ string \eqref{eq:AdS-string-classical}
and (in the strong coupling limit) the non-dyonic giant magnon \eqref{eq:E-expansion}
much simpler than this one.

\subsection{Subleading corrections\label{sub:Subleading-corrections-elementary-magnon}}

Consider first the new sum prescription, for which we need to evaluate
\eqref{eq:deltaE-subleading-new}. Using \eqref{eq:Alg-curve-Small-mag},
and taking the non-dyonic case $X^{\pm}=e^{\pm ip/2}$, we can write
the following pieces of that integral:\begin{align}
F_{\mathrm{heavy}}^{+}=\sum_{\substack{ij\\
\mathrm{heavy}}
}(-1)^{F_{ij}}e^{-i\left[q_{i}(x)-q_{j}(x)\right]} & =e^{-2i\frac{\alpha x}{x^{2}-1}}\frac{(x+1)\left(e^{ip/2}-1\right)\left(e^{ip/2}(3+x)-(3x+1)\right)}{\left(x-e^{ip/2}\right)^{2}}\displaybreak[0]\nonumber \\
F_{\mathrm{light}}^{+}=\sum_{\substack{ij\\
\mathrm{light}}
}(-1)^{F_{ij}}e^{-i\left[q_{i}(x)-q_{j}(x)\right]} & =e^{-i\frac{\alpha x}{x^{2}-1}}\frac{4(x+1)\left(e^{ip/2}-1\right)}{x-e^{ip/2}}.\label{eq:F-light-nondyonic}\end{align}
The contribution from the heavy modes will clearly be subleading to
that from the light modes, so we need only consider the latter. The
resulting expression for $\delta E^{F,1}$ agrees with that in \cite{Shenderovich:2008bs}.
This can be integrated using the saddle point at $x=i$ to give \cite{Bombardelli:2008qd}:%
\footnote{Recall that $\alpha/2=\Delta/\sqrt{2\lambda}=\Delta/4g$.%
} \begin{equation}
\delta E_{\mathrm{new}}^{F,1}=e^{-\Delta/\sqrt{2\lambda}}\sqrt{\frac{2\,\sqrt{2\lambda}}{\pi\:\Delta}}\left(\frac{\cos\frac{p}{2}}{1-\sin\frac{p}{2}}-1\right).\label{eq:delta-E-F1-small}\end{equation}
This term was also calculated by \cite{Bombardelli:2008qd} using
the L\"{u}scher method, obtaining exactly the same answer. That calculation
is really in terms of $h(\lambda)$ not $\lambda$; however this comparison
tests only the leading order part of \eqref{eq:two-expansions-of-h},
$h(\lambda)=\sqrt{\lambda/2}$, and tells us nothing about $c$. 

Now consider the other sums:
\begin{itemize}
\item For the old sum, the change in \eqref{eq:deltaE-subleading-old} is
that while we integrate the light modes at $\epsilon$, for the heavy
modes we use $2\epsilon$. At this order we need only note that this
will not change the fact that the heavy modes are subleading, and
so the answer is the same. 
\item For the alternative new sum, the change is that for the heavy modes
\eqref{eq:deltaE-subleading-BL} has instead \[
\sum_{\substack{ij\\
\mathrm{heavy}}
}(-1)^{F_{ij}}e^{-i\frac{q_{i}(x)-q_{j}(x)}{2}}=e^{-i\frac{\alpha x}{x^{2}-1}}\frac{3(x-e^{ip/2})+(x\, e^{ip/2}-1)-4\sqrt{x-e^{ip/2}}\sqrt{x\, e^{ip/2}-1}}{x-e^{ip/2}}.\]
This is now of the same order as the light modes, and so must be included.
Doing so changes the result to \begin{equation}
\delta E_{\mathrm{new}'}^{F,1}=e^{-\Delta/\sqrt{2\lambda}}\sqrt{\frac{2\,\sqrt{2\lambda}}{\pi\:\Delta}}\left(\frac{2+4\cot\frac{p}{4}}{\cot\frac{p}{4}-1}-4\sqrt{\frac{\cot\frac{p}{4}+1}{\cot\frac{p}{4}-1}}\right)\label{eq:delta-E-F1-alternative-new-sum}\end{equation}
which clearly disagrees with the L\"{u}scher result. 
\end{itemize}
To summarise, we obtain the desired subleading correction using either
the old or the new sum prescription. However the alternative new sum
of \cite{Bandres:2009kw} gives a mismatching result.

\subsubsection*{Dyonic case}

It is trivial to generalise the above results to the dyonic case.
The integrand in \eqref{eq:F-light-nondyonic} becomes \begin{flalign*}
F_{\mathrm{light}}^{\pm}(x) & =\sum_{\substack{ij\\
\mathrm{light}}
}(-1)^{F_{ij}}e^{\mp i(q_{i}-q_{j})}\partial_{x}\Omega_{45}(x)\\
 & =e^{\mp i\alpha\frac{x}{x^{2}-1}}\left\{ e^{\pm ip/2}\left[\left(\frac{x-X^{\mp}}{x-X^{\pm}}\right)^{2}+\left(\frac{x-1/X^{\pm}}{x-1/X^{\mp}}\right)^{2}\right]\left(\frac{x-X^{\pm}}{x-X^{\mp}}+\frac{x-1/X^{\mp}}{x-1/X^{\pm}}\right)\right.\\
 & \qquad\qquad\qquad\left.-2\left(\frac{x-X^{+}}{x-X^{-}}\right)\left(\frac{x-1/X^{+}}{x-1/X^{-}}\right)-2\left(\frac{x-X^{-}}{x-X^{+}}\right)\left(\frac{x-1/X^{-}}{x-1/X^{+}}\right)\right\} \partial_{x}\Omega_{45}(x).\end{flalign*}
The result of integrating this (using the saddle point at $x=i$)
is hardly more compact, so we write simply\begin{equation}
\delta E^{F,1}=\frac{1}{\sqrt{4\pi\alpha}}F_{\mathrm{light}}^{+}(i)\label{eq:dyonic-F-term}\end{equation}
 for both the old and the new sums.

\subsection{Sub-subleading terms\label{sub:Sub-subleading-terms}}

We can see from \eqref{eq:F-light-nondyonic} that the heavy modes
first contribute at order $(e^{-\Delta/\sqrt{2\lambda}})^{2}$. The
full correction at this order will also include the contribution of
the light modes from the $m=2$ term in \eqref{eq:dE-F-all-m}.%
\footnote{In equation \eqref{eq:a_mn-expansion} below, this term $\delta E^{F,2}$
is the term containing $a_{2,0}$. Note that while $\delta E^{F}$
in \eqref{eq:dE-F-all-m} adds up all the terms $a_{m,0}$ in \eqref{eq:a_mn-expansion},
the heavy modes in \eqref{eq:dE-F-all-m}'s term $m$ contribute to
$a_{2m,0}$.%
} The integrand of this term contains (in the non-dyonic limit) \[
F_{\mathrm{light}}^{+2}=\sum_{\substack{ij\\
\mathrm{light}}
}(-1)^{F_{ij}}e^{-2i\left[q_{i}(x)-q_{j}(x)\right]}=e^{-2i\frac{\alpha x}{x^{2}-1}}\frac{4(x^{2}-1)\left(e^{ip}-1\right)}{(x-e^{ip/2})^{2}}.\]
Putting these two contributions together, the correction for the new
sum is given by: \begin{align}
\delta E_{\mathrm{new}}^{F,2} & =-\frac{1}{8\pi i}\sum_{\pm}\int_{\mathbb{U}_{\pm}}dx\sum_{ij}(-1)^{F_{ij}}\left[q'_{i}(x)-q'_{j}(x)\right]\begin{cases}
\left(\pm i2e^{\mp2i(q_{i}-q_{j})}\right)\Omega_{ij}(x), & (i,j)\mbox{ heavy}\\
\left(\pm i2e^{\mp i(q_{i}-q_{j})}\right)\Omega_{ij}(x), & (i,j)\mbox{ light}\end{cases}\displaybreak[0]\nonumber \\
 & =-\frac{1}{2\pi i}\int_{\mathbb{U}_{+}}dx\Biggl[\sum_{\substack{ij\\
\mathrm{light}}
}(-1)^{F_{ij}}\frac{1}{2}e^{-2i(q_{i}-q_{j})}\partial_{x}\Omega_{ij}(x)+\sum_{\substack{ij\\
\mathrm{heavy}}
}(-1)^{F_{ij}}e^{-i(q_{i}-q_{j})}\partial_{x}\Omega_{ij}(x)\Biggr]\displaybreak[0]\nonumber \\
 & =-\frac{1}{2\pi i}\int_{\mathbb{U}_{+}}dx\left[\frac{1}{2}\; F_{\mathrm{light}}^{+2}\;\Omega_{45}'(x)+F_{\mathrm{heavy}}^{+}\;2\Omega_{45}'(x)\right]\nonumber \\
 & =e^{-2\Delta/\sqrt{2\lambda}}\;2\sqrt{\frac{\sqrt{2\lambda}}{\Delta\:\pi}}\left(\frac{\cos\frac{p}{2}-1}{\sin\frac{p}{2}-1}\right).\label{F2}\end{align}

For the old sum, the essential point to notice is that the heavy and
light terms above each lead to a finite contribution, and thus we
may take the limits $\epsilon\to0$ individually. This removes the
only distinction between the new and the old sum here, and so we obtain
the same result:

\[
\delta E_{\mathrm{old}}^{F,2}=\delta E_{\mathrm{new}}^{F,2}.\]
We would not expect $\delta E^{F,2}$ to depend on the value of $c$,
since like the subleading term $\delta E^{F,1}$ it is the first term
in a series in $1/\sqrt{\lambda}$. The extra power of $e^{-\Delta/\sqrt{2\lambda}}$
to be sub-subleading makes this a different series, not the second
term in the series. 

Finally, for the alternative new sum, we will again get a different
result, just as for the $\delta E^{F,1}$ term in \eqref{eq:delta-E-F1-alternative-new-sum}
above.

\section{Conclusions\label{sec:Conclusions}}

All calculations of the $AdS_{4}\times CP^{3}$ interpolating function
$h(\lambda)$ work by comparing an expansion in $h(\lambda)$, coming
from some integrable structure, to an expansion in $\lambda$, coming
from either gauge theory (expanding about $\lambda=0$) or string
theory (about $\lambda=\infty$). Such comparisons include:
\begin{itemize}
\item The gauge theory calculations of \cite{Minahan:2009aq,Minahan:2009wg}
use the exact dispersion relation and draw Feynman diagrams up to
four loops, order $\lambda^{4}$. 
\item For $AdS_{3}$ spinning strings an expansion of the Bethe equations
\cite{Gromov:2008qe} is compared to a semiclassical calculation using
either the worldsheet sigma-model \cite{McLoughlin:2008ms,Alday:2008ut,Krishnan:2008zs}
or algebraic curves \cite{Gromov:2008fy}.
\item The leading ($J=\infty$) corrections for giant magnons in this paper
(and in \cite{Shenderovich:2008bs}) are computed using the algebraic
curve, and compared with the exact dispersion relation \eqref{eq:disp-rel-ABJM}.
\item For finite-$J$ corrections we can compare instead to the L\"{u}scher
formulae, which take as input the all-loop S-matrix of \cite{Ahn:2008aa}.
This S-matrix is constructed to agree with the all-loop Bethe ansatz,
and thus similarly contains $h(\lambda)$. 
\end{itemize}
In all of these cases, analogous calculations have been done in $AdS_{5}\times S^{5}$,
and always agree with the trivial interpolating function $h(\lambda)=\lambda$.
Indeed, such comparisons essentially constitute the experimental evidence
for the simple form of the interpolating function for this theory
\cite{Hofman:2006xt,Gaiotto:2008cg}. There is also an argument \cite{Berenstein:2009qd}
that S-duality fixes the form of $h(\lambda)$ exactly; this is not
expected to exist in the $AdS_{4}\times CP^{3}$ case.

Higher-order perturbative checks have also been done, and a strong-coupling
result which would be particularly valuable to have here is the two-loop
comparison of spinning strings in $AdS_{5}$ with the Bethe ansatz
\cite{Giombi:2010fa,Gromov:2008en,Roiban:2007ju}. At two loops, the
$AdS_{4}$ radius is expected to receive corrections \cite{Bergman:2009zh},%
\footnote{However there are no corrections to $R$ at one loop, see also \cite{McLoughlin:2008he}
for another argument. Therefore this issue, of quantum corrections
to \eqref{eq:relation-R4-lambda-N/k}, does not overlap with the present
issue of scheme-dependence of one-loop corrections $\delta E$ and
thus of the coefficient $c$. \\
\indent In the $AdS_{5}\times S^{5}$ case, the topic of $\alpha'$
corrections to $R$ (the lack thereof) was studied in \cite{Banks:1998nr,Kallosh:1998zx}
and \cite{Mazzucato:2009fv}.%
} so one would potentially learn about these in addition to the next
term in $h(\lambda)$. 

Like the Bethe ansatz which they generalise, the recently proposed
TBA and Y-system descriptions \cite{Gromov:2009tv,Bombardelli:2009xz,Gromov:2009at}
are in terms of $h(\lambda)$ rather than $\lambda$. It is in order
to be able to translate new results from such descriptions back into
the original string- or field-theory language that we need to know
about $h(\lambda)$.

\subsection{Results at $J=\infty$\label{sub:conclusions-infinite-J}}

We calculated the one-loop energy correction for infinite-$J$ giant
magnons using three different summation prescriptions, which we called
old, new \cite{Gromov:2008fy}, and alternative new \cite{Bandres:2009kw}.
We find that one can either
\begin{itemize}
\item use the old sum prescription and set $c=-\log(2)/2\pi$, or
\item use either of the new sum prescriptions and set $c=0$.
\end{itemize}
This is precisely the same scheme-dependence as was seen for spinning
strings in $AdS$. We obtain it however from strings moving only in
$CP^{3}$, whose one-loop corrections are finite, rather than growing
as $\log S$, and are functions of two variables ($p$ and $Q$, encoded
in $X^{\pm}$). 

\bigskip 

On a technical level, this scheme-dependence comes from a logarithmic
divergence in the sum over heavy or light modes alone, which cancels
between them. The contributions of heavy and light modes are the two
terms in \eqref{eq:small-mag-DeltaE-old}:\begin{align*}
\delta E & =\lim_{\epsilon\to0}\left[\vphantom{\frac{a}{b}}-L(\epsilon_{\mathrm{heavy}})+L(\epsilon)\right]\\
 & \quad\mbox{where }L(\epsilon)=\frac{1}{i\pi}\Big(\frac{X^{+}-X^{-}}{1+X^{+}X^{-}}\Big)\log\epsilon+\mbox{finite terms},\end{align*}
and $\epsilon_{\mathrm{heavy}}=\epsilon$ for either new sum, $\epsilon_{\mathrm{heavy}}=2\epsilon$
for the old sum.%
\footnote{Here $\left|x\right|>1+\epsilon$ is the cutoff in the spectral plane.
In terms of the mode sum cutoff $N$, it is $\epsilon=\alpha/4\pi N$.%
}

A similar cancellation of logarithmic divergences between heavy and
light modes lies behind the finite results of the $AdS_{3}$ spinning
string calculations of \cite{McLoughlin:2008ms,Alday:2008ut,Krishnan:2008zs,McLoughlin:2008he}
(using the old sum) and \cite{Gromov:2008fy} (new sum), even though
these papers display only the combined, finite, results.\bigskip 

The heavy modes are something of a puzzle, since the Bethe equations
refer only the light modes (4 bosons and 4 fermions) while the string
theory treats all 10 dimensions alike. In the formalism used here,
each heavy mode is constructed off-shell as the sum of two light modes,
\eqref{eq:generate-lots-inv}. However we note that this is not true
for the \emph{on-shell} modes whose frequencies enter into the energy
correction. 

It has been argued that when loop corrections are taken into account,
the heavy states dissolve into the continuum of two-particle states
\cite{Zarembo:2009au}, see also \cite{Sundin:2009zu,Astolfi:2009qh,Bykov:2010tv}.
However the fact that they are not stable particles in the interacting
theory does not imply that they should be omitted from the path integral,
and indeed the present calculation requires that they be included
in order to obtain a finite result. 

\bigskip 

Finally, we observe that it is the old sum which comes closest to
imposing a physical cutoff, treating all modes on an equal footing.
The frequencies $\omega=\Omega(x)$ computed here are frequencies
with respect to physical time. Unlike the mode number (or worse, the
position in the spectral plane) this is a local quantity on the worldsheet.
If we explicitly choose the same cutoff for heavy and light modes,
by setting $\Omega_{\mathrm{heavy}}(1+\epsilon_{\mathrm{heavy}})=\Omega_{\mathrm{light}}(1+\epsilon)=\Lambda$,
then the vacuum's frequencies \eqref{eq:Omega-vac} lead us to\[
\epsilon_{\mathrm{heavy}}=2\epsilon+\mathcal{O}(\epsilon^{2}).\]
Using instead the giant magnon's frequencies \eqref{eq:Omega-small}
gives no change at this order. And from the point of view of the calculation
of $\delta E$ in \eqref{eq:small-mag-DeltaE-old}, this physical
condition is equivalent to the old sum, \eqref{eq:deltaE-old}.

\subsection{Finite-$J$ effects\label{sub:conclusions-finite-J}}

Following \cite{Gromov:2008ie,Gromov:2008ec} we can summarise the
complete energy of a giant magnon, including the various finite $J$
(thus finite $\Delta$) corrections, as follows:\begin{equation}
E=\negthickspace\negthickspace\sum_{m,n=0,1,2\ldots}\negthickspace\negthickspace a_{m,n}\left(e^{-\Delta/\sqrt{2\lambda}}\right)^{m}\left(e^{-2\Delta/E}\right)^{n}.\label{eq:a_mn-expansion}\end{equation}
Each of the coefficients $a_{m,n}$ is a series in $1/\sqrt{\lambda}$,
and the leading corrections discussed above are part of $a_{0,0}=E_{\mathrm{class}}+\delta E+o(1/\sqrt{\lambda})$.

The coefficients $a_{m,0}$ are classically zero. Calculating $a_{1,0}$
at one-loop (order $\sqrt{\lambda}^{0}$) following \cite{Gromov:2008ie}
we see no difference between the old and new sums, and find agreement
with the L\"{u}scher method calculation of \cite{Bombardelli:2008qd}.%
\footnote{The dyonic $a_{1,0}$ term given in \eqref{eq:dyonic-F-term} has
now been confirmed by a bound-state L\"{u}scher F-term calculation
in a recent paper \cite{Ahn:2010eg}. %
} (There these are referred to as F-terms, and arise from virtual particles
travelling full circle around the worldsheet.) However for the alternative
new sum of \cite{Bandres:2009kw}, we find a disagreement. In this
case both heavy and light modes contribute. For the old and new sums,
$a_{1,0}$ depends only on the light modes, with heavy modes first
entering in $a_{2,0}$, which we also calculate, \eqref{F2}.

The coefficient $a_{0,1}$ contains the classical (order $\sqrt{\lambda}$)
corrections to the magnon's energy, of they type studied by \cite{Arutyunov:2006gs,Astolfi:2007uz,Okamura:2006zv,Hatsuda:2008gd}%
\footnote{For the corresponding solutions in $RP^{2}$ and $CP^{1}$ (inside
$CP^{3}$) see \cite{Grignani:2008te,Lee:2008ui,Ahn:2008wd} and \cite{Abbott:2008qd}.%
} and, using algebraic curves, by \cite{Minahan:2008re,Lukowski:2008eq,Abbott:2009um}.
Its one-loop part was calculated for the $AdS_{5}\times S^{5}$ case
by \cite{Gromov:2008ec}, who found agreement with the subleading
L\"{u}scher $\mu$-term calculation of \cite{Janik:2007wt}. The
analogue of their calculation is useful to us here because, like $a_{0,0}$,
the one-loop corrections give us the \emph{second} term in the series
in $1/\sqrt{\lambda}$, and so we can potentially learn about $c$. 

In order to calculate the relevant quantum corrections, we need to
start with the algebraic curve for a classical finite-$J$ giant magnon.
This, and the need to keep various terms we ignored before, adds considerable
complication \cite{Gromov:2008ec}. In addition not all of the L\"{u}scher
terms one would like to compare to are known. Thus far we can report
that: 
\begin{itemize}
\item The leading bound-state $\mu$-term matches perfectly with the classical
algebraic curve result of \cite{Abbott:2009um} for one dyonic elementary
magnon.%
\footnote{Note that we believe the leading \emph{single-magnon} $\mu$-term
calculations of \cite{Lukowski:2008eq,Bombardelli:2008qd} to be incorrect,
since they give zero rather than the AFZ result expected for a non-dyonic
giant magnon \cite{Arutyunov:2006gs,Abbott:2009um}.%
} 
\item The subleading $\mu$-terms of \cite{Bombardelli:2008qd} for the
$RP^{3}$ magnon can be recovered from the one-loop algebraic curve
by calculating $a_{0,1}$ using the new sum. 
\item When calculating $a_{0,1}$ using the old sum \eqref{eq:deltaE-old},
it has a linear divergence in the cutoff $N$. 
\end{itemize}
These and other related calculations are the material of a forthcoming
paper.

\subsection*{Acknowledgements}

From the inception of this project, we thank an anonymous referee
of the paper \cite{Bombardelli:2008qd}, and the organisers of the
Potsdam IGST conference, July 2009. Along the way, we thank C. Ahn,
D. Fioravanti, M. Kim and S. Minwalla for discussions. 

For hospitality while working on this, MCA thanks Wits (Johannesburg)
and IST (Lisbon), IA thanks TIFR (Mumbai), and DB thanks IEU (Seoul). 

IA was supported in part by the Funda\c{c}\~{a}o para a Ci\^{e}ncia
e a Tecnologia (FCT / Portugal). DB was supported by WCU grant No.
R32-2008-000-101300, and by University PRIN 2007JHLPEZ `\emph{Fisica
Statistica dei Sistemi Fortemente Correlati all'Equilibrio e Fuori
Equilibrio: Risultati Esatti e Metodi di Teoria dei Campi}'.

\appendix

\section{The Classical Algebraic Curve\label{sec:The-Classical-Algebraic-Curve}}

For completeness we give here some relevant properties. The $AdS_{4}\times CP^{3}$
case was first studied by \cite{Gromov:2008bz}, drawing on past work
on $AdS_{5}\times S^{5}$ by \cite{Kazakov:2004qf,Kazakov:2004nh,Beisert:2004ag,SchaferNameki:2004ik,Beisert:2005bm,Beisert:2005di}
among others. 

The monodromy matrix is defined from the Lax connection $J(x)$ by
\[
\Omega(x)=P\, e^{\oint d\sigma J_{\sigma}(x)}.\]
Here we integrate once around the worldsheet $(\sigma,\tau)$. The
connection depends on an arbitrary complex number $x$ called the
spectral parameter, and since it is flat (for all $x$) the eigenvalues
of $\Omega$ are independent of the path used. We write these as \[
\eig\Omega(x)=\left\{ \vphantom{\frac{a}{a}}e^{i\hat{p}_{1}},e^{i\hat{p}_{2}},e^{i\hat{p}_{3}},e^{i\hat{p}_{4}},e^{i\tilde{p}_{1}},e^{i\tilde{p}_{2}},e^{i\tilde{p}_{3}},e^{i\tilde{p}_{4}}\right\} \]
and call to the eight functions $\tilde{p}_{i}$ ($CP$) and $\hat{p}_{i}$
($AdS$) `quasi-momenta'. In order to make the $OSp(2,2|6)$ symmetry
explicit, we will work not with $p_{i}$ but instead with ten new
quasi-momenta $q_{i}$ defined \cite{Gromov:2008bz} \[
\Big(q_{1},q_{2},q_{3},q_{4},q_{5}\Big)=\frac{1}{2}\Big(\hat{p}_{1}+\hat{p}_{2},\hat{p}_{1}-\hat{p}_{2},\tilde{p}_{1}+\tilde{p}_{2},-\tilde{p}_{2}-\tilde{p}_{4},\tilde{p}_{1}+\tilde{p}_{4}\Big)\]
and $\big(q_{6},q_{7},q_{8},q_{9},q_{10}\big)=\big(-q_{5},-q_{4},-q_{3},-q_{2},-q_{1}\big).$

These functions define a 10-sheeted Riemann surface. It need not however
be continuous across branch cuts, so long as $\eig\mathrm{\Omega(x)}$
is continuous: when a cut $C_{ij}$ connects sheets $i$ and $j$,
we must have $q_{i}^{+}-q_{j}^{-}=2\pi n$ when $x\in C_{ij}$. There
are two additional constraints: 
\begin{itemize}
\item First, the Virasoro constraints lead to synchronised poles at $x=\pm1$:
\begin{align*}
\big(q_{1},q_{2},q_{3},q_{4},q_{5}\big) & =\frac{\alpha_{+}}{(x-1)}\big(1,1,1,1,0\big)+\mathcal{O}(x-1)^{0}\\
 & =\frac{\alpha_{-}}{(x+1)}\big(1,1,1,1,0\big)+\mathcal{O}(x+1)^{0}.\end{align*}

\item Second, the curve has the following inversion symmetries: \begin{align}
q_{1}(\tfrac{1}{x}) & =-q_{2}(x)\nonumber \\
q_{3}(\tfrac{1}{x}) & =2\pi m-q_{4}(x)\label{eq:class-inversion-cond}\\
q_{5}(\tfrac{1}{x}) & =q_{5}(x).\nonumber \end{align}
This $m\in\mathbb{Z}$ is the winding number.
\end{itemize}
The string's charges are determined by the asymptotic behaviour as
$x\to\infty$: \begin{equation}
\left(\begin{array}{c}
q_{1}\\
q_{2}\\
q_{3}\\
q_{4}\\
q_{5}\end{array}\right)=\left(\begin{array}{c}
0\\
0\\
-p/2\\
-p/2\\
0\end{array}\right)+\frac{1}{2gx}\left(\begin{array}{c}
\Delta+S\\
\Delta-S\\
(J+Q)/2\\
(J-Q)/2\\
J_{3}\end{array}\right)+\mathcal{O}\Big(\frac{1}{x^{2}}\Big).\label{eq:asympt-curve-charge}\end{equation}
 The `twists' $p/2$ are the same as we added to the solution \eqref{eq:Alg-curve-Small-mag}
to allow for nonzero momentum. In \cite{Abbott:2009um} we instead
allowed non-integer $m$; this however will get the $F$-terms of
section \ref{sub:Subleading-corrections-elementary-magnon} wrong. 

For each square-root branch cut $C_{ij}$, we define the filling fraction
as\begin{equation}
S_{ij}=\frac{g}{i\pi}\oint_{C_{ij}}dx\left(1-\frac{1}{x^{2}}\right)q_{i}(x).\label{eq:integral-filling-fraction}\end{equation}
The new poles which we add when studying fluctuations are very short
branch cuts; this is why they connect two sheets $i$ and $j$. The
residue $\alpha(y)$ is set by the condition that they are each exactly
one fluctuation: $S_{ij}=1$.

\section{Corrections for the Big Giant Magnon\label{sec:Corrections-Big}}

The big giant magnon is a two-parameter one-angular-momentum solution,
which was known in the algebraic curve \cite{Shenderovich:2008bs}
before being constructed in the $\sigma$-model \cite{Hollowood:2009tw,Kalousios:2009mp,Suzuki:2009sc,Hatsuda:2009pc}.
It should be thought of as consisting of two elementary magnons in
a particular orientation \cite{Hollowood:2009sc,Hatsuda:2009pc}.%
\footnote{These two elementary magnons have the same worldsheet velocity \cite{Hollowood:2009sc}.
Superpositions of two elementary magnons having different velocies
are instead scattering solutions \cite{Hatsuda:2009pc}. See also
\cite{Kalousios:2010ne} for more than two magnons.%
} 

The big magnon is described by the algebraic curve\begin{align*}
q_{1}(x)=q_{2}(x) & =\dfrac{\alpha x}{x^{2}-1}\\
q_{3}(x)=q_{4}(x) & =\dfrac{\alpha x}{x^{2}-1}+G(x)+G(0)-G(\tfrac{1}{x})-p\\
q_{5}(x) & =0\end{align*}
with $G(x)$ as in \eqref{eq:defn-Gmag}. As for the $RP^{3}$ case,
we now adopt conventions in which $2p$ is the total momentum.

\subsection{Off-shell frequencies}

For the $(1,5)$ polarisation, we use this ansatz:\begin{align*}
\delta q_{1} & =\frac{\alpha(y)}{x-y}+\frac{a_{+}}{x+1}+\frac{a_{-}}{x-1} & \delta q_{2} & =-\delta q_{1}(\tfrac{1}{x})\\
\delta q_{4} & =\frac{a_{+}}{x+1}+\frac{a_{-}}{x-1}+M(x)+M(0)-M(\tfrac{1}{x}) & \delta q_{3} & =-\delta q_{4}(\tfrac{1}{x})\\
\delta q_{5} & =\frac{\alpha(y)}{x-y}+\frac{\alpha(y)}{\frac{1}{x}-y}+\frac{\alpha(y)}{y}.\end{align*}
This clearly has the correct new poles and satisfies the inversion
symmetries, and the changes in the residues of the poles at $x=\pm1$
are all the same.%
\footnote{The fluctuation $\delta q_{i}$ given in \cite{Shenderovich:2008bs}
uses the following terms instead: \begin{align*}
\frac{\alpha(x)}{x-y} & =\frac{\alpha(y)}{x-y}\;+\frac{1}{4g}\sum_{\pm}\frac{1}{(1\pm y)(x\pm1)}\\
\frac{\alpha(x)/x}{x-X^{+}} & =\frac{\alpha(X^{+})/X^{+}}{x-X^{+}}\;-\frac{1}{4g}\sum_{\pm}\frac{1}{(X^{+}\pm1)(x\pm1)}\end{align*}
These are chosen to automatically give the right behaviour at infinity,
and it is then the equations at $x=\pm1$ which fix $\delta\Delta$.%
} Imposing the conditions at infinity now fixes $a_{\pm}$ and $A^{\pm}$,
and we get\[
\Omega_{15}(y)=\delta\Delta=\frac{1}{y^{2}-1}\left(1-y\frac{X^{+}+X^{-}}{1+X^{+}X^{-}}\right).\]

For the $(4,5)$ polarisation, \begin{align*}
\delta q_{1} & =\frac{a}{x+1}+\frac{a}{x-1} & \delta q_{2} & =-\delta q_{1}(\tfrac{1}{x})\\
\delta q_{4} & =-\frac{\alpha(y)}{x-y}+\frac{a}{x+1}+\frac{a}{x-1}+M(x)+M(0)-M(\tfrac{1}{x}) & \delta q_{3} & =-\delta q_{4}(\tfrac{1}{x})\\
\delta q_{5} & =\frac{\alpha(y)}{x-y}+\frac{\alpha(y)}{\frac{1}{x}-y}+\frac{\alpha(y)}{y}\end{align*}
leads to\[
\Omega_{45}(y)=\frac{1}{y^{2}-1}\left(1-y\frac{X^{+}+X^{-}}{1+X^{+}X^{-}}\right).\]

We can then find all the other frequencies, and as before: \begin{equation}
\Omega_{ij}(y)=\begin{cases}
\Omega_{45}(y), & (i,j)\mbox{ light}\\
2\Omega_{45}(y), & (i,j)\mbox{ heavy}.\end{cases}\label{eq:Omega-big}\end{equation}
This is exactly as in \cite{Shenderovich:2008bs}, except for notation.

\subsection{Energy corrections}

Note that identity \eqref{eq:sum-qq-1234-identity} still holds, and
with \eqref{eq:Omega-big} implies that for the new sum, we get \[
\delta E_{\mathrm{new}}=0.\]
This is the result of \cite{Shenderovich:2008bs}. Despite pre-dating
\cite{Gromov:2008fy}, this paper uses an integral like \eqref{eq:deltaE-new-GM},
and thus has implicitly adopted the new sum \eqref{eq:sum-new-GM}.

For the old sum, \eqref{eq:sum-naiive}, a similar calculation to
the one we did for the elementary giant magnon leads to \[
\delta E_{\mathrm{old}}=-2\frac{\log2}{\pi}\sin\frac{p}{2}\]
twice what we got for the elementary magnon, equation \eqref{eq:small-DeltaE-old-nondyonic},
and thus consistent with the same value of $c$.

\section{Corrections for the $RP^{3}$ Magnon\label{sec:Corrections-RP3}}

The $RP^{3}$ giant magnon, an embedding of Dorey's $S^{3}$ dyonic
giant magnon, is described in the algebraic curve by\begin{align*}
q_{1}(x)=q_{2}(x) & =\dfrac{\alpha x}{x^{2}-1}\\
q_{3}(x) & =\dfrac{\alpha x}{x^{2}-1}+2G(0)-2G(\tfrac{1}{x})-p\\
q_{4}(x) & =\dfrac{\alpha x}{x^{2}-1}+2G(x)-p\\
q_{5}(x) & =0.\end{align*}
This is a superposition of two elementary magnons (one `\emph{u}'
and one `\emph{v}' in \cite{Abbott:2009um}). Here $p$ is the momentum
of each of the elementary magnons, so that the total momentum is $2p$.%
\footnote{This allows us to still write $X^{\pm}=r\, e^{\pm ip/2}$ or $p=-i\log(X^{+}/X^{-})$.
In our previous paper \cite{Abbott:2009um} we instead defined $p$
as the total momentum. However, $\Delta$, $J$, and $Q$ are still
the total charges. %
} The dispersion relation is\begin{align}
E & =\sqrt{\frac{Q^{2}}{4}+16h(\lambda)^{2}\sin^{2}\frac{p}{2}}\label{eq:disp-rel-RP3}\\
 & =8g\sin\frac{p}{2}\qquad\mbox{when }Q=2\ll\lambda=8g^{2}.\nonumber \end{align}
In the non-dyonic limit (and at strong coupling) this is simply twice
that of the elementary magnon. 

The calculation of this one-loop correction is very similar to that
for one elementary magnon, so we state results here without showing
any detail. Using the new sum, we obtain\[
\delta E_{\mathrm{new}}=0\]
and using the old sum \[
\delta E_{\mathrm{old}}=-2\frac{\log2}{\pi}\sin\frac{p}{2}\]
exactly twice that for the elementary magnon, and thus consistent
with the same value of $c$. 

Finally the subleading correction for both old and new sums is\begin{align}
\delta E^{F,1} & =\frac{1}{\sqrt{\alpha\pi}}e^{-\alpha/2}\left[e^{ip}\frac{X^{+2}(X^{-2}+1)^{2}+X^{-2}(X^{+2}+1)^{2}}{X^{+2}(X^{+}-i)^{2}(X^{-}+i)^{2}}-2\right]\nonumber \\
 & =\frac{-4}{\sqrt{\alpha\pi}}e^{-\alpha/2}\frac{\sin\frac{p}{2}}{\sin\frac{p}{2}-1}\qquad\mbox{when }Q\ll\sqrt{\lambda}.\label{eq:F-term-RP3}\end{align}
In the non-dyonic limit this correction matches the L\"{u}sher F-term
calculated by \cite{Bombardelli:2008qd}. In this limit the $RP^{3}$
magnon and the big magnon co-incide, and the same integral was also
obtained for this term by \cite{Shenderovich:2008bs}.

\section{Conventions and the Vacuum\label{sub:Vacuum-BL}}

When defining the frequency $\Omega(y)$ from $\delta\Delta$, the
paper \cite{Bandres:2009kw} writes, instead of our \eqref{eq:delta-Delta-and-sum-Omega},
\begin{equation}
\Omega(y)=\delta\Delta+\sum_{AdS\,\mathrm{modes}}N^{ij}+\frac{1}{2}\sum_{\mathrm{fermions}}N^{ij}.\label{eq:Omega-dD-with-BL-shifts}\end{equation}
This change (from our conventions) cancels out of either of the new
sums, but not out of the old sum. As a result that paper finds that
$\delta E_{\mathrm{old}}=\infty$ for the vacuum. Since the same shifts
apply to any soliton solution too, they will cancel out of any normalised
energy correction $\delta E-\delta E_{\mathrm{vac}}$. Our conventions
have the advantage of producing a much simpler set of frequencies
\eqref{eq:Omega-vac}. The conventions of \cite{Gromov:2008bz} agree
with those of \cite{Bandres:2009kw}, since they obtain the same frequency
shifts although without writing a formula like \eqref{eq:Omega-dD-with-BL-shifts}. 

We observe that our conventions produce off-shell frequencies which
vanish as the new pole is taken to infinity: $\Omega(y\to\infty)=0$. 

Similar calculations of the same vacuum frequencies have been done
from the worldsheet perspective, either directly \cite{Bandres:2009kw}
or using the Penrose limit \cite{Nishioka:2008gz,Gaiotto:2008cg,Grignani:2008is},
obtaining various other constant shifts. We summarise these in table
\ref{tab:vacuum-shifts}. 

Note that all of these are only constant shifts added to the frequencies.
What the paper \cite{Mikhaylov:2010ib} discusses is half-integer
shifts of $n$, which are much more subtle. The conclusion there was
that one has to be very careful to get these right for fermions in
the worldsheet calculation.

\begin{table}
\begin{centering}
\begin{small}\begin{tabular}{l|ccc|cccc|}
 &  & \hspace*{-10mm}Algebraic Curves:\hspace*{-10mm} &  &  & Worldsheet:\hspace*{-15mm} &  & \tabularnewline
 & This paper & G\&V \cite{Gromov:2008bz} & B\&L \cite{Bandres:2009kw} & N\&T \cite{Nishioka:2008gz} & GGY \cite{Gaiotto:2008cg} & GHO \cite{Grignani:2008is} & B\&L \cite{Bandres:2009kw}\tabularnewline
\hline
$AdS$ bosons &  &  &  &  &  &  & \tabularnewline
~~~~~~heavy $\times3$ & $-1$ & 0 & 0 & 0 & 0 & 0 & 0\tabularnewline
Fermions &  &  &  &  &  &  & \tabularnewline
~~~~~~heavy $\times4$ & $-1$ & $-1/2$ & $-1/2$ & 0 & 0 &  & $\pm1/2$\tabularnewline
~~~~~~light $\times4$ & $-1/2$ & 0 & 0 & 0 & 0 &  & 0\tabularnewline
$CP$ bosons &  &  &  &  &  &  & \tabularnewline
~~~~~~heavy $\times1$ & $-1$ & $-1$ & $-1$ & 0 & 0 & 0 & 0\tabularnewline
~~~~~~light $\times4$ & $-1/2$ & $-1/2$ & $-1/2$ & 0 & 0 & $\pm1/2$ & $\pm1/2$\tabularnewline
\hline 
Total (weighted)  & 0 & $-1$ & $-1$ & 0 & 0 &  & 0\tabularnewline
\end{tabular}\end{small}
\par\end{centering}

\caption{Constant shifts of the vacuum's fluctuation frequencies. The `unshifted'
frequencies are the square root terms in \eqref{eq:w_n-vac} above,
and the total of course includes multiplicity and counts fermions
with a minus.\label{tab:vacuum-shifts}}

\end{table}

\section{Momentum Conservation and Level Matching\label{sec:Momentum-Conservation-etc}}

\subsection{The vacuum}

When constructing the perturbations $\delta q$ for the vacuum (point
particle) solution, the paper \cite{Bandres:2009kw} used a pair of
new poles at $\pm y$, and calculate the total $\delta\Delta=\Omega(y)+\Omega(-y)$.
This construction is clearly blind to any terms odd in $y$. However
it is justified in this case, since on-shell we have $x_{-n}^{ij}=-x_{n}^{ij}$
for all $ij,n$, and every sum $\delta E$ contains $\omega_{n}+\omega_{-n}$,
so such terms cannot affect the result. 

Using a pair of excitations $x_{\pm n}^{ij}$ is also sufficient to
satisfy the level matching condition \eqref{eq:level-matching}, although
it will not be the only way to do so. The paper \cite{Gromov:2008bz}
states that they always use a pair of poles $x_{\pm n}^{ij}$ for
this reason, and for the vacuum case they study this is equivalent
to using a pair at $\pm y$. 

Another way to construct $\Omega(y)$ is to use just one pole but
allow some change in the momentum: for the $(4,5)$ polarisation,
we would use this ansatz:\begin{align}
\delta q_{1} & =\sum_{\pm}\frac{a_{\pm}}{x\pm1} & \delta q_{2} & =-\delta q_{1}(\tfrac{1}{x})\nonumber \\
\delta q_{4} & =-\frac{\alpha(y)}{x-y}+\sum_{\pm}\frac{a_{\pm}}{x\pm1} & \delta q_{3} & =-\delta q_{4}(\tfrac{1}{x})+\delta p\nonumber \\
\delta q_{5} & =\frac{\alpha(y)}{x-y}+\frac{\alpha(y)}{\frac{1}{x}-y}+\frac{\alpha(y)}{y}.\label{eq:dq-vac-with-momentum}\end{align}
This leads to the same frequency as before, \[
\Omega_{45}(y)=\delta\Delta=\frac{1}{y^{2}-1}\]
as well as momentum \begin{equation}
\delta p=\frac{\alpha(y)}{y}=\frac{1}{2g}\frac{y}{y^{2}-1}.\label{eq:vac-dp}\end{equation}
It is clear that when considering two poles at $\pm y$, the total
$\delta p$ will be zero again.

When we construct a heavy fluctuation like this, such as the $(3,7)$
mode, we will get $\delta p=2\alpha(y)/y$. Alternatively recall that
we constructed heavy fluctuations in \eqref{eq:generate-lots-inv}
by adding two light fluctuations, and the $\delta p$ will similarly
add up.

\subsection{Giant magnons}

We can repeat our analysis of the giant magnon allowing $\delta p\neq0$,
in the same way as for the vacuum: change the ansatz \eqref{eq:small-dq-ansatz}
to have $\delta q_{3}(x)=-\delta q_{4}(\tfrac{1}{x})+\delta p$. We
find (writing just the non-dyonic case)\begin{equation}
\Omega_{45}(y)=\left[\frac{1}{y^{2}-1}-\frac{y}{y^{2}-1}\cos\frac{p}{2}\right]+2g\,\delta p\,\cos\frac{p}{2}.\label{eq:O45-with-momentum}\end{equation}
The first two terms (in square brackets) are the terms appearing in
\eqref{eq:Omega-small}. Notice that if we take $\delta p=\alpha(y)/y$
(as for the vacuum) then the new third term here has the same form
as the second term --- in fact they cancel. 

This is a nice demonstration of the argument for the giant magnon's
off-shell frequency $\Omega(y)$ given by \cite{Gromov:2008ie}. They
say that the first term is the energy of the excitation, while the
second term comes from the fact that the perturbation carries some
momentum $\delta p$, and so if total momentum is conserved, the magnon's
momentum must change to compensate. We can write this as \[
\Omega(y)=E_{\mathrm{excitation}}-\delta p\frac{\partial E_{\mathrm{magnon}}}{\partial p}\]
and we recall that $\partial E/\partial p=2g\cos\frac{p}{2}$ for
the non-dyonic case. 

We can describe such an excitation using $X^{\pm}$ near to $y$:
solving $Q=1$ in \eqref{eq:Q-in-gXpXm} we find \[
X^{\pm}=y\pm\frac{i}{4g}\frac{y^{2}}{y^{2}-1}\]
which leads to the same momentum as \eqref{eq:vac-dp} \[
\delta p=-i\log\frac{X^{+}}{X^{-}}=\frac{1}{2g}\frac{y}{y^{2}-1}.\]

Now consider the effect of the term in $\Omega(y)$ arising from momentum
conservation on our calculation of the one-loop correction $\delta E$.
If we drop the second term from all $\Omega_{ij}(y)$, there will
be no change in the new sum \eqref{eq:small-mag-DeltaE-new}. But
there is a change in the old sum \eqref{eq:small-mag-DeltaE-old},
which becomes \[
\delta E_{\mathrm{old}}=-2\frac{\log2}{2\pi}\frac{1}{\sin\frac{p}{2}}\]
This is not a term which $c$ could produce in \eqref{eq:E-expansion}. 

Had we constructed $\delta q(x)$ starting with a pair of new poles
at $\pm y$, we would not have obtained the second term in $\Omega_{ij}(y)$,
since it is odd in $y$. For the elementary magnon it is clear that
we may not do this, as \eqref{eq:pole-position} always involves $G_{\mathrm{mag}}(x)$
which has no simple behaviour under $x\to-x$. But in the case of
the big magnon, for the (1,5) polarisation, we may have been tempted:
the classical curve's $q_{1}(x)$ and $a_{5}(x)$ are identical to
those of the vacuum. For the $(4,5)$ polarisation, clearly we cannot.
Dropping this second term from $\Omega_{15}(y)$ but not from $\Omega_{45}(y)$,
and then blindly using section \ref{sub:Off-shell-method}'s formulae
to generate all the rest, we obtain a divergent correction $\delta E$.
And our error is that for instance $\Omega_{18}(y)$ has been built
using $\Omega_{15}(y)$, and thus a pair of new poles $\pm y$, but
for this polarisation the on-shell pole positions are not simply related,
$x_{\pm n}^{18}\neq\pm y$.

\subsection{Spinning strings}

The paper \cite{Bandres:2009kw} studies quantum corrections for spinning
strings with two equal angular momenta. When constructing the fluctuation
$\delta q(x)$, it uses a pair of poles at $\pm y$. This is justified
for both the $(1,5)$ and the $(4,5)$ polarisations, just as it was
for the vacuum. 

But it is not justified for all the other polarisations: not all of
the on-shell pole positions come in pairs $x_{\pm n}^{ij}=\pm y$.
For example the $(1,8)$ polarisation does not have this property.
Nevertheless $\Omega_{18}(y)$ is constructed from $\Omega_{15}(y)$
and $\Omega_{45}(y)$. 

Attempting to find a way to construct $\delta q(x)$ without using
this assumption, we tried allowing both $\delta p$ and for the two
endpoints of the square-root cut to move independently. This appears
to lead to a valid fluctuation, which adds to the result of \cite{Bandres:2009kw}
the following term in $\Omega_{45}(y)$:\begin{equation}
i\frac{m}{\mathcal{J}}\frac{K(1/y)}{K(1)}\frac{y}{y^{2}-1}.\end{equation}
The change in the momentum is \begin{equation}
\delta p=\frac{\alpha(y)}{y}(1-i)\frac{K(1/y)}{\mathcal{J}}.\end{equation}
Here we use that paper's notation: $K(x)=\sqrt{\mathcal{J}^{2}+m^{2}x^{2}/4}$
is the branch cut term, with $\mathcal{J}$ the angular momentum and
$m$ the winding. 

It's not entirely clear what to make of these new terms. We have not
tried to work out whether they affect the energy corrections for which
agreement was found with worldsheet results.

\begin{small}\bibliographystyle{my-JHEP-3}
\addcontentsline{toc}{section}{\refname}\bibliography{/Users/me/Documents/Papers/complete-library-processed,complete-library-processed}

\end{small}

\end{document}